\documentclass[11pt]{article}

\usepackage{graphicx}
\usepackage{subfig}
\usepackage{amsmath}
 \usepackage{amsfonts}
\usepackage{epsfig}

\newcommand{\be}{\begin{equation}}
\newcommand{\ee}{\end{equation}}
\newcommand{\bes}{\begin{equation*}}
\newcommand{\ees}{\end{equation*}}
\newcommand{\beqn}{\begin{eqnarray}}
\newcommand{\eeqn}{\end{eqnarray}}
\newcommand{\beqns}{\begin{eqnarray*}}
\newcommand{\eeqns}{\end{eqnarray*}}

\newcommand{\lkr}{\left(}
\newcommand{\lkv}{\left[}
\newcommand{\rkv}{\right]}
\newcommand{\rkr}{\right)}
\newcommand{\lfi}{\left\{}
\newcommand{\rfi}{\right\}}

\newcommand{\fr}[1]{(\ref{#1})}

\newcommand{\EE}{\ensuremath{{\mathbb E}}}

\newcommand{\II}{\ensuremath{{\mathbb I}}}

\newcommand{\PP}{\ensuremath{{\mathbb P}}}

\newcommand{\RR}{\ensuremath{{\mathbb R}}}

\newcommand{\Span}{\mbox{Span}}

\newcommand{\Var}{\mbox{Var}}

\newcommand{\diag}{\mbox{diag}}
\newcommand{\supp}{\mbox{supp}}

\newcommand{\Tr}{\mbox{Tr}}
\newcommand{\proj}{\mbox{proj}}

\newtheorem{theorem}{Theorem}
\newtheorem{lemma}{Lemma}

\newcommand{\bb}{\mathbf{b}}
\newcommand{\bc}{\mathbf{c}}
\newcommand{\bd}{\mathbf{d}}

\newcommand{\bh}{\mathbf{h}}

\newcommand{\bt}{\mathbf{t}}
\newcommand{\bu}{\mathbf{u}}

\newcommand{\bz}{\mathbf{z}}

\newcommand{\bA}{\mathbf{A}}

\newcommand{\bI}{\mathbf{I}}
\newcommand{\bQ}{\mathbf{Q}}

\newcommand{\bW}{\mathbf{W}}

\newcommand{\bY}{\mathbf{Y}}

\newcommand{\lam}{\lambda}
\newcommand{\del}{\delta}
\newcommand{\sig}{\sigma}
\newcommand{\ph}{\phi}
\newcommand{\Up}{\Upsilon}

\newcommand{\eps}{\varepsilon}

\newcommand{\lamin}{\lam_{\min}}
\newcommand{\lamax}{\lam_{\max}}

 \newcommand{\te}{\theta}
 \newcommand{\bte}{\mbox{\mathversion{bold}$\te$}}
 \newcommand{\bnu}{\mbox{\mathversion{bold}$\nu$}}
\newcommand{\bxi}{\mbox{\mathversion{bold}$\xi$}}
\newcommand{\boeta}{\mbox{\mathversion{bold}$\eta$}}
\newcommand{\bbe}{\mbox{\mathversion{bold}$\beta$}}

\newcommand{\bobeta}{\mbox{\mathversion{bold}$\beta$}}
 
\newcommand{\bphi}{\mbox{\mathversion{bold}$\phi$}}

\newcommand{\tpsi}{\widetilde{\psi}}

\newcommand{\bPhi}{\mbox{\mathversion{bold}$\Phi$}}
\newcommand{\bUp}{\mbox{\mathversion{bold}$\Up$}}

\newcommand{\calD}{{\mathcal{D}}}
\newcommand{\calG}{{\mathcal G}}

\newcommand{\calL}{{\mathcal{L}}}
\newcommand{\calP}{{\mathcal{P}}}

\newcommand{\Jc}{{J^c}}

\graphicspath{%
    {converted_graphics/}
    {/}
}
\begin{document}

\title
{\bf Estimation of delta-contaminated density of the random intensity   of Poisson data}

\author{
 {\em Daniela De Canditiis},\\
               Istituto  per le Applicazioni
               del Calcolo "M. Picone",
               CNR Rome, Italy \\
                \\
          {\em Marianna Pensky},\\
          Department of Mathematics,
          University of Central Florida \\
         }

\date{}

\bibliographystyle{plain}
\maketitle

\begin{abstract}
In the present paper, we constructed an estimator of a delta contaminated mixing density function $g(\lam)$ 
of the intensity $\lambda$ of the Poisson distribution.
The estimator is based on an expansion of the continuous portion $g_0(\lambda)$ of the unknown pdf over an overcomplete dictionary
with the recovery of the coefficients  obtained as   solution of an optimization problem with Lasso penalty. In order to  
apply Lasso technique in the, so called, prediction setting where it requires virtually no assumptions on dictionary 
and, moreover, to ensure fast convergence of Lasso estimator, we use a novel formulation of the optimization problem
based on inversion of the dictionary elements. 
The total estimator of the  delta contaminated mixing pdf is obtained using a two-stage iterative procedure.

We formulate conditions on the dictionary and the unknown mixing density that yield a sharp oracle inequality 
for the norm of the difference between $g_0 (\lambda)$ and its estimator and, thus, obtain a   smaller error  than
in a minimax setting. Numerical simulations and comparisons with the  Laguerre functions based estimator 
recently constructed by Comte and Genon-Catalot  (2015) also show advantages of our procedure. 
At last, we apply the technique  developed in the paper to estimation of 
a delta contaminated mixing density of the Poisson intensity of the Saturn's rings data.\\
\vspace{3mm} {\bf  Keywords}: {Mixing density, Poisson distribution, empirical Bayes, Lasso penalty }\\
\vspace{3mm}{\bf AMS (2000) Subject Classification}: {Primary  62G07, 62C12. Secondary 62P35}

\end{abstract}

\section {Introduction}
\label{sec:intro}
\setcounter{equation}{0}

Poisson-distributed data appear  in many contexts. In the last two decades a large amount of effort was 
spent on recovering the mean function in the Poisson regression model.
%
In this set up, one observes independent Poisson variables  $Y_1, \cdots, Y_n$ where $Y_i$  with  respective means 
$\lam_i = f(i/n)$, $i=1, \cdots, n$. Here, $f$ is the function of interest which is assumed to   exhibit  some degree of smoothness.
The difficulty in estimating $f$ on the basis of Poisson data stems from the fact that the variances of the Poisson 
random variables are equal to their means and, hence, do  not remain constant as $f$ changes its values.
Estimation techniques are either   based on  variance stabilizing transforms  (Brown {\it et al.}  (2010),
Fryzlewicz and Nason (2004)), wavelets (Antoniadis and Sapatinas (2004), Besbeas {\it et al.} (2004),
Harmany {\it et al.}  (2012)), Haar frames (Hirakawa and Wolfe  (2012)) or Bayesian methods
(Kolaczyk (1999) and Timmermann and Nowak (1999)). The case of estimating Poisson intensity 
in the presence of missing data was studied in He {\it et al.}  (2005).

The fact that the variance of a Poisson random variable is equal to its mean serves as a common and reliable test 
that data in question are indeed Poisson distributed. However, in many practical situations, although each of the 
data value $Y_i \sim Poisson (\lam_i)$, $i=1, \cdots, n$, the overall data do not have Poisson distribution. 
This is due to the fact that consecutive values of $\lam_i$ are so different from each other that $f$ is not really 
a function. In this case, in order  to account for the extra-variance, it is usually reasonable to assume 
that $\lam$  itself is a random variable with an unknown probability density function $g$ which needs to be estimated.

In particular, below we consider the following problem. Let $\lam_i$, $i=1, \cdots, n$, be independent random 
variables that are  not observable  and have an unknown pdf $g(\lam)$. One observes   variables $Y_i|\lam_i \sim Poisson(\lam_i)$,
$i=1, \cdots,n$, that, given $\lam_i$,  are independent. Our objective is to estimate  $g (\lam)$, 
the so   called   {\it mixing density}, on the basis of  observations  $Y_1, \cdots, Y_n$. 
Here, $g$ can be viewed as the prior density of the parameter $\lam$,
so that the model above reduces to an empirical Bayes model where the prior has to be estimated from data.

Estimation of the prior density of the parameter of the Poisson distribution 
has been considered by several authors. For example,  
Lambert  and Tierney (1984) suggested non-parametric maximum likelihood estimator, 
 Walter (1985)  and Walter and Hamedani (1991) studied estimators based on Laguerre 
polynomials, Zhang (1995) considered smoothing kernel estimators and Herngartner (1997)
investigated Fourier series based estimators of $g$. All papers listed above provided the 
upper bounds for the mean integrated   squared error (MISE);  Zhang (1995) and Herngartner (1997) also presented the lower bounds   
for the MISE over smoothness classes. The common feature of all these estimators is that the convergence rates 
are  very low. In particular, if $n \to \infty$, both Zhang (1995)  and Herngartner (1997)
obtained convergence rates of the form $(\ln n/\ln \ln n)^{-2\nu}$ where $\nu$ is the parameter of the 
smoothness class to which $g$ belongs.   The latter seem to imply that there is no hope for accurate estimation 
of the mixing density $g$ unless the sample sizes are extremely high.
On a more positive note, in  a recent paper, Comte and Genon-Catalot (2015)
considered an estimator of $g$ based on expansion of $g$ over the orthonormal Laguerre basis. They showed that if Laguerre 
coefficients of $g$ decrease exponentially, then the resulting estimator 
has convergence rates that are polynomial in $n$ and provided some examples where this happens.  
Moreover, they proposed a penalty for controlling the number of terms in the expansion and provided   
oracle inequalities for the estimators of $g$ under various scenarios.

The low convergence rates for the prior density of Poisson parameter are due to the fact that its recovery 
constitutes a particular case of an ill-posed linear inverse problem. Indeed, let $L^2[0,\infty)$ and $\ell^2$ 
be the Hilbert spaces of, respectively, square integrable functions on $[0,\infty)$ and square integrable sequences.
Denote the probability that $Y=l$, $l=0,1, \cdots$, by  $P(l) = \PP(Y=l)$.  Then,  introducing a linear operator 
$Q: L^2[0,\infty) \rightarrow \ell^2$, we can present  $g(\lam)$ as the solution of the following equation
\be \label{main_eq}
(Qg) (l) = \int_0^\infty \frac{\lam^l \, e^{-\lam}}{l!} g(\lam)\, d\lam = P(l), \quad l=0,1, \cdots
\ee
Since exact values of the probabilities $P(l)$ are unknown, they can be estimated by relative frequencies  $\nu_l$,
so the problem of recovering $g$ appears as an ill-posed linear inverse problem with the right-hand side measured with error. 
Solution of equation \fr{main_eq} is particularly challenging since the $g$ is a  function  of a real argument while $P$ is the 
infinite-dimensional vector.

In the last decade, a great deal of effort was spent on recovery of an unknown function   
in regression setting from its noisy observations using overcomplete  dictionaries. 
In particular, if the dictionary is large enough and the function of interest has a 
sparse representation in this dictionary, then it can be recovered with a much better precision than   
when  it is expanded over an orthonormal basis. Lasso and its versions 
(see e.g. B\"uhlmann and van de Geer (2011) and references therein)
allow one to identify the dictionary elements that guarantee efficient estimation  of the unknown 
regression function. The advantage of this approach is that the estimation error is controlled by the, 
so called, oracle inequalities that provide   upper bounds for the risk for the particular function 
that is estimated rather than   convergence rates designed for the ``worst case scenario'' 
of the minimax setting. In addition, if the function of interest can be represented via a linear combination of 
just few dictionary elements, then  one can prove that it can be estimated with nearly parametric 
error given certain assumptions on the dictionary hold.

In the present paper, we  extend  this idea to the case of estimating a mixing density $g$ 
on the basis of $Y_1, \cdots, Y_n$.  However, there is an intrinsic difficulty arising from the fact that the 
 problem above is an ill-posed  inverse problem. Currently, one can justify convergence of a Lasso estimator
only if  stringent assumptions on the dictionary, the, so called, compatibility conditions, are satisfied. 
In regression set up, as long as compatibility conditions hold, one can prove that 
Lasso estimator is nearly optimal.
Regrettably, while compatibility conditions may be satisfied for the functions   in the  original dictionary,
they usually do not hold for their images  due to contraction imposed by the   operator $Q$.
In the present paper, we show how   to  circumvent this difficulty and apply Lasso methodology to 
estimation of $g$.  We formulate conditions on the dictionary and the unknown mixing density that yield a sharp oracle inequality 
for the norm of the difference between $g  (\lam)$ and its estimator and, thus, result in a   smaller error  than
in a minimax setting.  
Numerical simulations and comparisons with the  Laguerre functions based estimator 
recently constructed by Comte and Genon-Catalot  (2015) also show advantages of our procedure.

Our study is motivated by analysis of the astronomical data, in particular, the photon counts $Y_i, i=1, \cdots, n$
that come from sets of observations   of stellar occultations   
recorded by the Cassini UVIS high speed photometer at different radial points  on the Saturn's ring plane.
It is well known that Saturn ring is comprised of particles of various sizes,   each on its own orbit about Saturn. 
With no outside influences, these photon counts should follow  the Poisson distribution, however, 
obstructions imposed by the particles in the ring cause photon counts  distribution to deviate from Poisson.
 The latter is due to the fact that  although, for each $i=1, \cdots, n$,
the photon counts $Y_i \sim Poisson (\lam_i)$, the values of $\lam_i$, $i=1, \cdots, n$, are
extremely  varied and, specifically, are best described as random variables with the unknown underlying pdf $g(\lam)$.

In addition, if a ring region contains a significant proportion of large  particles,
those particles can completely block out  the light leading to zero photon counts.
For this reason, we assume that the unknown pdf $g$ is delta-contaminated, 
i.e., it is a combination of an unknown mass $\pi_0$ at zero and a 
continuous part, so that $g(\lam)$ can be written as 
\be \label{pdf_g}
g (\lam) = \pi_0 \del(\lam) + (1 - \pi_0) g_0(\lam)
\ee
where  $g_0(\lam)$ is an unknown pdf and $\del(\lam)$ is the Dirac delta function such that, 
for any integrable function $f$ one has $\int f(x) \delta(x) d x = f(0)$. 
Models of the type \fr{pdf_g} also appear in other applied settings (see, e.g., Lord {\it et al.} (2005)).
However, to the best of our knowledge, we are the first ones to estimate the delta-contaminated density of the
intensity parameter of the Poisson distribution. 
In this setting, we also obtain a sharp oracle inequality 
for the norm of the difference between $g_0 (\lam)$ and its estimator.
We also derive convergence rates for the estimator $\widehat{\pi_0}$  of the mass $\pi_0$ at zero.
The estimator has  also been  successfully applied to recovery of   delta-contaminated densities of the intensities 
$\lambda$ for various sub-regions of the Saturn's rings.

Finally, we should remark on several other advantages  of the approach presented in the paper. 
First, although in the paper we are using the gamma dictionary, the technique can be applied with any type of dictionary functions
since it is based on a numerical inversion of dictionary elements. Moreover, the method can be used 
even if the underlying conditional distribution is different from Poisson. 
The estimator exhibits no boundary effects and performs well in simulations delivering small errors.
Moreover, since we apply Tikhonov regularization for recovering inverse images of the dictionary elements,
our estimator can be viewed as a version of an elastic net estimator (\cite{zou}).

The rest of the paper is organized as follows. 
Sections~\ref{sec:lasso_est}~and~\ref{sec:implement} present, respectively, 
the method  and the algorithm for construction of an estimator of the unknown density function, 
while Section~\ref{sec:error} studies its convergence properties.
 Section~\ref{sec:simulation} investigates precision of   the 
estimators developed in the paper via numerical simulations with synthetic data.
Section~\ref{sec:real data} provides application of the technique proposed   in the paper 
to the occultation data for the Saturn's rings.
Finally, Section~\ref{sec:proofs} contains the proofs of the statements presented in the paper.


\section{The Lasso estimator of the mixing density} 
\label{sec:lasso_est}
\setcounter{equation}{0}

In what follows, we assume that $g_0(\lam)$ in \fr{pdf_g} can be well approximated by a dictionary 
that consists  of gamma pdfs  
\begin{equation} \label{def_phi}
\phi_k(\lambda)=\gamma(\lambda; a_k,b_k)=\frac{\lambda^{a_k-1} \exp(-\lambda/b_k)}{b_k^{a_k} \Gamma(a_k)}, 
\quad k=1, \cdots, p.
\end{equation}
This is a natural assumption since, for any fixed $b_k=b$ and $a_k = 1,2, \cdots,$ the linear span of 
$\phi_k, k=1, \cdots$, coincides with the space $L^2 (0, \infty)$, so that a linear combination of $\phi_k, k=1, \cdots, p$,
with large $p$ approximates any square integrable function with a small error.  
Indeed, for a fixed $b_k =b$ and $a_k = 1,2,3, \cdots$, this dictionary contains linear combinations of 
the Laguerre functions and, hence, its span approximates $L^2 [0,\infty)$ space. 
On the other hand, using a variety of scales $b_k$ allows one 
to accurately represent a function of interest with many fewer terms.

Using this dictionary, we estimate $g$ by 
\be \label{hatg}
\hat{g} (\lam) = \hat{\pi}_0 \del(\lam) + (1 - \hat{\pi}_0)\,  \sum_{k=1}^p \widehat{c_k} \phi_k(\lam), 
\ee
applying a two-step procedure.
If the estimator $\hat{\pi}_0$ were already constructed, coefficients   $c_k$, $k=1, \cdots, p$, 
could be chosen, so to minimize the squared $L^2$-norm   
\begin{align}  \label{continuous01}
 \| g  -\hat{g} \|^2_2  = 
  \| g  -\hat{\pi}_0 \delta \|^2_2 + 
(1-\hat{\pi}_0)^2\ \left\| \sum_{k=1}^p c_k \phi_k \right\|^2_2 
- 2(1-\hat{\pi}_0)\, \sum_{k=1}^p c_k \langle g - \hat{\pi}_0 \delta, \phi_k   \rangle. 
\end{align}
The first term in formula \fr{continuous01} does not depend on coefficients $c_k$ while the second term 
is completely known. In order to estimate the last term, note that 
$\langle g - \hat{\pi}_0 \delta, \phi_k   \rangle = \langle g, \phi_k \rangle - \hat{\pi}_0 \phi_k(0)$.
Moreover,  if we found functions $\chi_k \in \ell^2$ such that 
\begin{equation} \label{def_psi}
 (Q^* \chi_k) (\lam) =  \sum_{i=0}^\infty   \frac{e^{-\lambda}  \lambda^i}{i!}\, \chi_k(i)=\phi_k(\lambda),  \quad \forall \lambda \in (0, +\infty),
\end{equation}
 then, it is easy to check that 
\begin{eqnarray}  
\langle g, \phi_k  \rangle & = &  \int_0^{+\infty}  g(\lambda)  \sum_{i=0}^\infty   \frac{e^{-\lambda}  \lambda^i}{i!}\, \chi_k(i) d \lambda  
= \sum_{i=0}^\infty \chi_k(i) \int_0^{+\infty}  g(\lambda) \frac{e^{-\lambda}  \lambda^i}{i!}  d \lambda \nonumber\\
& = & \sum_{i=0}^\infty \chi_k(i) P(i)  = \EE   \chi_k (Y).  \label{second}
\end{eqnarray}
Here, $P(l)$ is  the marginal probability function  
\be \label{marg_prob_fun}
P(l) = \PP(Y=l) = \pi_0\, \II(l=0)+(1-\pi_0) \sum_{k=1}^p c_k U_k(l),  \quad l=0,1,2,\cdots
\ee
where $\II(l=0)$ is the indicator that $l=0$. 
Hence, $\langle g, \phi_k  \rangle $ can be estimated by 
\be \label{third_term}
\widehat{ \langle g, \phi_k   \rangle} =   n^{-1} \sum_{i=1}^n \chi_k(Y_i) =  
\sum_{l =0}^\infty \chi_k(l) \nu_l = \langle \chi_k, \nu \rangle, \quad k=1,\ldots,p,
\ee
where 
\be \label{rel_freq}
\nu_l = n^{-1} \, \sum_{i=1}^n \II(Y_i=l), \quad l=0,1, \cdots 
\ee
are the relative frequencies of $Y=l$ and $\II(A)$ is the indicator function of a set $A$.


There is an obstacle to carrying out estimation above. Indeed, for some values of $a_k$ and $b_k$ in formula 
\fr{def_phi}, solutions $\chi_k (Y)$ of equations \fr{def_psi}  may not have finite  variances  or variances may be too high. 
In order to stabilize the variance we use Tikhonov regularization. In particular, we replace solution 
$\chi_k = (Q^*)^{-1} \phi_k$ of equation \fr{def_psi} by solution $\tpsi_{k, \zeta_k}$ of equation
\be \label{psikzeta}
(Q Q^* + \zeta_k I) \tpsi_{k, \zeta_k} = Q \phi_k, \quad \zeta_k>0,
\ee
where operators $Q$ and $Q^*$ are defined in \fr{main_eq} and \fr{def_psi}, respectively, and $I$ is the identity operator,
so that, for any $f$,  
$$
(Q Q^*  f)(j) = \sum_{l=0}^\infty {j+l \choose l} 2^{-(j+l+1)} f(l), \quad j=0,1,\ldots
$$
Observe that $\Var  [\tpsi_{k, \zeta_k} (Y)]$ is a decreasing function of $\zeta_k$ while the squared bias 
$(\EE \tpsi_{k, \zeta_k} -  \langle g, \phi_k  \rangle )^2$ is an increasing function of $\zeta_k$. 
Denote $\widehat{\zeta_k}$ the unique solution of the following equation
\be \label{hatzeta}
\frac{1}{n} \Var [\tpsi_{k, \widehat{\zeta_k}} (Y)] = 
\left( \EE \tpsi_{k, \widehat{\zeta_k}} -  \langle g, \phi_k  \rangle \right)^2
\ee
and replace $\chi_k(Y)$ in \fr{third_term} by 
\be \label{psik_sigmak}
\psi_k(Y) = \tpsi_{k, \widehat{\zeta_k}} (Y) \quad \mbox{with} \quad \sig^2_k = \Var [\psi_k(Y)]. 
\ee

In order to identify the correct subset of dictionary functions $\phi_k$, we introduce a weighted Lasso penalty.  
In particular, the vector of coefficients $\hat{\bc}$ with components $\hat{c}_k, k=1, \cdots, p$, 
can be recovered as a solution of the following optimization problem 
\be \label{lasso_for_c}
\hat{\bc} = \underset{\bc}{\operatorname{argmin}}~
 \left\{(1-\hat{\pi}_0)^2\ \left\| \sum_{k=1}^p c_k \phi_k \right\|^2_2  
 - 2(1-\hat{\pi}_0)\, \sum_{k=1}^p c_k  \left[ \langle \psi_k, \nu \rangle - \hat{\pi}_0\, \phi_k(0) \right] + 
\alpha_c \sum_{k=1}^p \sigma_k |c_k| \right\}.
\ee
Here,  $\sum_{k=1}^p \sigma_k |c_k|$ is the weighted Lasso penalty and $\alpha_c$ is the penalty parameter.

Now, consider the problem of estimating the weight $\pi_0$ when coefficients $c_k$, $k=1, \cdots, p$, are known. 
Denote
\begin{equation} \label{def_Uk}
U_k(l)= \int_0^{+\infty} \frac{e^{-\lambda}  \lambda^l}{l!}\, \phi_k(\lambda) d \lambda=
\frac{\Gamma(l+a_k)}{\Gamma(a_k)\, l!} \  b_k^{l} (1+ b_k)^{-(l+a_k)} 
\end{equation}
and recall that  the marginal probability function is of the form \fr{marg_prob_fun}.
Hence,  up to the term that does not depend on $\pi_0$, 
given the data vector $\bY$ and the vector of coefficients $\bc$, the log-likelihood of $\pi_0$ can be written as 
\be \label{loglike}
\log L(\pi_0 | \bY, \bc) = n \, \nu_0 \, \log \left(\pi_0 + (1-\pi_0) \sum_{k=1}^p c_k U_k(0) \right) + n (1 - \nu_0) \log(1-\pi_0).
\ee
If  $\bu$ is a vector  with components $u_k = U_k(0)$, then  the   expression \fr{loglike} is maximized by 
\be \label{pi_0_MLE}
\hat{\pi}_0^{MLE}=   \frac{{\nu}_0 - \bc^T \bu}{1- \bc^T \bu}, 
\ee

In order to implement optimization procedure suggested above,
consider  matrix $\bPhi \in \RR^{p \times p}$ with  elements $\Phi_{l k} = \langle \phi_k, \phi_l \rangle$, 
$l,k = 1, \cdots, p$, and define vectors  $\bz$ and $\bxi$ in  $\RR^p$  with components  
\be \label{z_xi}
z_k=\phi_k(0), \quad
\xi_k = \langle \psi_k, \nu \rangle = \sum_{l=0}^{\infty} \psi_k(l)\, \nu_l = n^{-1}\ \sum_{i=1}^n \psi_k (Y_i).
\ee
Denote $\bte = (1-\hat{\pi}_0) \bc$ and re-write optimization problem   \fr{lasso_for_c}  
in terms of vector $\bte$  as 
$$
 \widehat{\bte} = \underset{\bte}{\operatorname{argmin}}  \left\{  \bte^T \bPhi \bte  
- 2 \bte^T (\bxi  - \hat{\pi}_0  \bz)   + \alpha \sum_{k=1}^p \sig_k |\te_k| \right\},
$$
where the penalty parameter $\alpha$ is related to $\alpha_c$ in \fr{lasso_for_c}  as 
$\alpha_c = (1 - \hat{\pi}_0) \alpha$. 
Introduce matrix $\bW$ such that $\bPhi = \bW^T \bW$ and vector 
\be \label{vec_eta}
\boeta =(\bW^T)^{+}(\bxi - \hat{\pi}_0   \bz)=  \bW \,(\bW^T \bW)^{-1} (\bxi - \hat{\pi}_0   \bz),
\ee 
where, for any matrix $\bA$, matrix $\bA^{+}$ is the Moore-Penrose inverse of $\bA$. 
Then, for a given value of $\hat{\pi}_0$, optimization problem \fr{lasso_for_c} appears as 
\be \label{lasso_for_theta}   
 \widehat{\bte} = \ \underset{\bte}{\operatorname{argmin}}  \left\{ \| \bW \bte - \boeta \|^2_2 + 
 \alpha  \sum_{k=1}^p \sig_k |\te_k| \right\}.
\ee
Now, we need to re-write an estimator for $\pi_0$ in terms of vector $\bte$. For this purpose, replace $\bc$ by  
$(1 - \hat{\pi}_0^{MLE})^{-1}\, \bte$ and solve equation \fr{pi_0_MLE} for $\hat{\pi}_0^{MLE}$ obtaining
$$
\hat{\pi}_0^{MLE} = \nu_0 - \bte^T \bu.
$$
Since $\pi_0 \geq 0$, we estimate $\pi_0$ by
\begin{equation} \label{estimatepio}
\hat{\pi}_0 = \max(0,\, \nu_0 - \bte^T \bu). 
\end{equation}

\section{Implementation of the  Lasso estimator} 
\label{sec:implement}
\setcounter{equation}{0}

Formulae \fr{lasso_for_theta} and \fr{estimatepio} suggest the following two-step optimization procedure.

\subsection*{Algorithm}
 
\begin{enumerate}
\item 
Evaluate sample frequencies $\nu_l$, $l=0,1,...$, given by formula \fr{rel_freq}.

\item 
Choose initial value $\hat{\pi}_0^{(0)}=\nu_0$ and  obtain the vector of 
coefficients $\widehat{\bte}^{(0)}$ minimizing \fr{lasso_for_theta}
 
\item 
Find new value of $\hat{\pi}_0$ using formula (\ref{estimatepio}) as 
$\hat{\pi}_0^{(j)} = \max[0,\, \nu_0 - (\widehat{\bte}^{(j-1)})^T \bu]$. 
Then, obtain new estimator $\widehat{\bte}^{(j)}$
of coefficients $\bte$ by minimizing \fr{lasso_for_theta}, with $\hat{\pi}_0 = \hat{\pi}_0^{(j)}$.
Repeat this step  for $j=1,2,...$ until one of the following stopping criteria is met: 
$$
(i)\  \hat{\pi}^{(j)}_0 =0; \quad \quad (ii)\  \|W \widehat{\bte}^{(j)}  - W \widehat{\bte}^{(j-1)} \|_2^2 < tol;\quad \quad
(iii)\  j > J_{\max}. 
$$ 
Here $tol$ and $J_{\max}$ are, respectively, the tolerance level and 
the maximal number of steps defined in advance.

\item 
Obtain  the estimator 
\be \label{hatg}
\hat{g}(\lambda)= \hat{\pi}_0 \delta(\lambda) +   \sum_{k=1}^p \hat{\te}_k \phi_k(\lambda).
\ee
\end{enumerate}

Note that the algorithm described above is significantly simplified if $\phi_k(0)=0$ for all $k=1, \ldots, p$. Indeed, in this case,
vector $\bz=0$, so that vector $\boeta$ in \fr{vec_eta} is independent of $\hat{\pi}_0$. In this case, one does not need iterative 
optimization. In particular, vector of coefficients  $\widehat{\bte}$ is recovered as solution of optimization problem 
\fr{lasso_for_theta} and $\hat{\pi}_0$ is constructed according to formula  \fr{estimatepio}. 
In the  present version of the paper, we considered this option. Indeed, in addition to computational convenience, choosing 
$\phi_k(0)=0$ for all $k=1, \ldots, p$, guarantees convergence of the Lasso estimator \fr{hatg}.

In order to implement Lasso estimator, for any $\zeta_k$, we need  to obtain a solution $\tpsi_{k, \zeta_k} $ of equation
\fr{psikzeta}. For his purpose,  we introduce a matrix  version $\bQ$ of  operator $Q$ 
in \fr{main_eq}. The elements of matrix  $\bQ$ are Poisson probabilities  $\bQ_{li}= e^{-x_i} (x_i)^l/(l!)$, where 
$x_i = i h$, $i=1,2...$, are the grid points at which we are going to recover $g(\lambda)$ and $h$ is the step size.
Introduce vectors $\bphi_k$ and $\bold{\tpsi_{k, \zeta_k} }$, $k=1,\ldots,p$,  with elements $\phi_k(x_i)$,  $i=1,2...$, and $\tpsi_k(l)$,
$l=0,1,...$, respectively. 
Then, for each $k=1,\ldots, p$, equation \fr{psikzeta} can be re-written as  
\begin{equation}  \label{tikhonov}
\bold{\tpsi_{k} } =(\bQ \bQ^T +  \zeta_k \bI)^{-1} \bQ \bphi_k,
\end{equation}
where $\bI$ is the identity matrix.  For the sake of finding $\widehat{\zeta_k}$  satisfying \fr{hatzeta},
we created a grid 
and chose $\widehat{\zeta_k}$ so that to minimize an absolute value of 
$ \widehat{\Var} [ \tpsi_{k, \zeta_k} (Y) ] - \left( \EE \tpsi_{k, \zeta_k} -  \langle g, \phi_k  \rangle \right)^2$
where $\widehat{\Var} [ \tpsi_{k, \zeta_k} (Y) ]$ is the sample variance of $\tpsi_{k, \zeta_k} (Y)$. 
After that, we evaluated $\psi_k(Y)$ in \fr{psik_sigmak} and replaced unknown variances $\sig^2_k$ in \fr{psik_sigmak}
by their sample counterparts.

\section {Convergence and estimation error}
 \label{sec:error}
 \setcounter{equation}{0}


Let $\hat{g} (\lam)$ be given by  \fr{hatg}.
In order to derive oracle inequalities for the error of $\hat{g} (\lam)$, 
we introduce the following notations.  
Let 
\be \label{flam}
f(\lam) = (1 - \pi_0 ) g_0 (\lam), \quad f_{\bt} = \sum_{j=1}^p t_j \phi_j. 
\ee
For any vector $\bt \in \RR^p$, denote  its $\ell^2$, $\ell^1$, $\ell^0$ and $\ell^\infty$ norms by, 
respectively,  $\| \bt\|_2$, $\| \bt\|_1$,  $\| \bt\|_0$ and $\| \bt\|_\infty$.
Similarly, for  any function $f$,   denote by $\| f \|_2$, $\| f\|_1$ and 
$\| f \|_{\infty}$ its $L^2$, $L^1$ and $L^{\infty}$ norms.
 Denote $\calP = \{1, \cdots, p\}$.   For any subset of indices  $J \subseteq \calP$, 
 subset $J^c$ is its complement in $\calP$ and  $|J|$ is its cardinality, so that $|\calP| =p$. 
Let  $ \calL_J = \Span \lfi \ph_j, \   j \in J \rfi$. 
If $J \subset \calP$ and $\bt \in \RR^p$,  then     $\bt_J \in \RR^{|J|}$ denotes   reduction of vector $\bt$ to
subset of indices $J$. Let $\bte_0$ be coefficients of the projection of $f = (1 - \pi_0) g_0$ on the 
linear span of the dictionary $\calL_\calP$, i.e., $f_{\bte_0} = \proj_{\calL_\calP} f$.
Let $J_0 = \supp (\bte_0)$.   
%
%
Denote by $\lamin (m)$ and $\lamax (m)$ the minimum and the maximum restricted eigenvalues
of matrix $\bPhi$ 
\be \label{eigrestrict}
\lamin (m) = \min_{\stackrel{\bt \in \RR^p}{\|\bt \|_0 \leq m}}\ \frac{\bt^T \bPhi \bt}{\| \bt \|_2^2}, 
\quad
\lamax (m) = \max_{\stackrel{\bt \in \RR^p}{\|\bt \|_0 \leq m}}\ \frac{\bt^T \bPhi \bt}{\| \bt \|_2^2}. 
\ee
Denote $\bUp = \diag(\sig_1, \cdots, \sig_p)$,
\be \label{Dset}
\calD (\mu, J)  = \lfi \bd \in \RR^p:\ \|(\bUp \bd)_{\Jc}\|_1 \leq \mu  \|(\bUp \bd)_{J}\|_1  \rfi, \quad \mu >1,
\ee
and  consider the set $\calG (C_{\sig})$ of subsets $J \subset \calP$  
\be \label{Gset}
\calG (C_{\sig}) = \lfi J \subset \calP:\  \max_{j \in J,\, j' \in \Jc} \frac{ \sig_j}{ \sig_{j'}} \leq C_{\sig}  \rfi.
\ee

It turns out that, as long as the sample size $n$ is large enough,
estimator $f_{\widehat{\bte}}$ is close  to $f$ with high probability,
with no additional assumptions. Indeed, the following statement holds.

\begin{theorem} \label{th:slow_Lasso}
Let  $\phi_k(0) = 0$, $k=1, \cdots, p$, and $n \geq N_0$ where
\be \label{N_cond}
N_0 = \frac{16}{9} (\tau +1) \log p \ \max_{1 \leq k \leq p} \lkv \frac{\|\psi_k\|_\infty^2}{\sig_k^2} \rkv.
\ee
 Then, for    any $\tau >0$  and any $\alpha \geq \alpha_0$, with probability at least
$1 - 2 p^{-\tau}$,  one has
\be \label{slowlas}
\| f_{\widehat{\bte}} - f \|_2^2 \leq \inf_{\bt } \lkv  \| f_{\bt } - f \|_2^2 + 4 \alpha   \sum_{j=1}^p \sig_j |t_j|  \rkv
\ee
where $\widehat{\bte}$ is the solution of optimization problem   \fr{lasso_for_theta} and
\be \label{alf0}
\alpha_0 = (2 \sqrt{(\tau +1) \log p} + 1)\, n^{-1/2} .
\ee
\end{theorem}


Theorem \ref{th:slow_Lasso} provides the, so called, slow Lasso rates. In order to obtain faster   
convergence rates and also to ensure that $\hat{\pi}_0$ is close to $\pi_0$ with high probability, 
we impose the following two conditions on the dictionary $\phi_k, k=1, \cdots, p,$ and the true function $f$.
The first  condition  needs to ensure  that   the dictionary $\lfi \ph_j,\ j \in \calP \rfi$ 
is incoherent and it can be warranted by  the following  
 assumption  introduced in  \cite{bickel_ritov_tsybakov}:
\\

\noindent
{\bf (A1)} \   \ For some $s$, $1 \leq s \leq p/2$, some $m \geq s$ 
such that $s+m \leq p$ and some constant $C_0$ one has
\be \label{a2b}
m\, \lamin (s+m) > C_0^2\, s\, \lamax (m),
\ee
where $\lamin (s+m)$   and $\lamax (m)$  are restricted eigenvalues of matrix $\bPhi$ defined in \fr{eigrestrict}.
\\

Observe  that, if $J \in  \calG (C_{\sig})$, condition $\bd \in \calD (\mu, J)$ implies that 
$\| \bd_{\Jc}\|_1 \leq \mu C_{\sig}  \|\bd_{J}\|_1$, so that Lemma 4.1. of Bickel {\it et al.} (2009)
yields that
\be \label{vartsm}
\vartheta (s,m, \mu, C_\sig) = \min_{\stackrel{J \in \calG (C_{\sig})}{|J| \leq s}}\, 
\min_{\stackrel{\bd \in \calD (\mu, J)}{\bd \neq 0}}\ \frac{\bd^T \bPhi \bd}{\| \bd_J \|_2^2} \geq
 \lamin(s+m) \lkr 1 - \frac{\mu C_{\sig} \sqrt{s \lamax(m)}}{\sqrt{m \lamin(s+m)}} \rkr^2 >0
\ee
provided Assumption {\bf (A1)} holds with $C_0 = \mu C_\sig$.


As a second condition, we assume that the true function $f$ in \fr{flam} is such that 
its  ``good'' approximation can be achieved using $J \in \calG (C_{\sig})$.
\\

\noindent
{\bf (A2)} For some $\mu>0$, $C_{\sig} >0$ and some $H_0>0$  one has 
\be \label{assA1}
 \widehat{J} = \arg\min_{J \subseteq \calP} \lkv   \| f - f_{\calL _J} \|_2^2 +  \frac{H_0 \log p}{\vartheta (s,m, \mu, C_\sig)}    
   \ \sum_{j \in J} \frac{\sig_j^2}{n} \rkv \in \calG (C_{\sig}). 
\ee
\\

Note that Assumption {\bf  A2} is natural and is similar to the usual assumptions
that $f$ is smooth and does not have  fast oscillating components. In the context of the ill-posed problems,
Assumption~{\bf  A2}  means that $f$ is not ``too hard'' to estimate.
Under Assumptions~{\bf  A1}~and~{\bf  A2}, one can prove the ``fast'' convergevce rates for $\hat{f}$ 
as well as obtain the error bounds for  $\hat{\pi}_0$.

\begin{theorem} \label{th:fast_Lasso}
Let  $\phi_k(0) = 0$, $k=1, \cdots, p$, $n \geq N_0$ where $N_0$ is defined in \fr{N_cond}.
Let $\alpha = \varpi \alpha_0$ where $\alpha_0$ is  defined in \fr{alf0}.
Let Assumptions  {\bf  A1 } and {\bf  A2 } hold with some $\mu$ and $C_\sig$,  $|\widehat{J}| \leq s$,  $C_0 = \mu C_\sig$  and
$H_0 \geq 2(1 + \varpi)^2 (4 \tau +5)$, where   $\varpi \geq  (\mu +1)/(\mu -1)$.
Then, with probability at least $1 - 2 p^{-\tau}$,  one has  
\be \label{f_error}
\| f_{\widehat{\bte}} - f \|_2^2   \leq \inf_{J \subseteq \calP} \lkv \| f - f_{\calL _J} \|_2^2 +   
\frac{4\, H_0 \log p}{\vartheta (s,m, \mu, C_\sig)}    
  \ \sum_{j \in J} \frac{\sig_j^2}{n} \rkv.
\ee  
Moreover, if $J_0  \in \calG (C_{\sig})$ and $|J_0| \leq  s$, then,  with probability at least $1 - 4 p^{-\tau}$,  one has
\be \label{pi_error}
(\hat{\pi}_0 - \pi_0)^2 \leq    \frac{4 H_0\, (1 + \mu C_{\sig})^2\,   \|\bu \|_\infty^2 \  s\, \log p}
{\vartheta^2 (s,m, \mu, C_\sig)}  \ \sum_{j \in J_0}  \frac{\sig_j^2}{n},  
\ee
where   $\bu$ is a vector  with components $u_k = U_k(0)$ defined in \fr{def_Uk}. 
\end{theorem}


\section {Numerical Simulations}
\label{sec:simulation}
\setcounter{equation}{0}

In order to evaluate the accuracy of the proposed estimator we carried out a simulation study where we tested 
performance of the proposed estimator under various scenarios. In order to assess precision of the estimator,
for each of the scenarios, we evaluated the relative integrated error of   $\hat{g}$ defined as 
\be \label{Del_g}
\Delta_g = \|g - \hat{g}\|_2^2  /\|g\|_2^2
\ee
where the norm was calculated over the grid $x_i = ih$ with $i=0,1, \ldots$, if $\hat{\pi}_0=\pi_0=0$ and
$i= 1,2, \ldots$, otherwise. 
In addition, we studied prediction properties of  $\widehat{g}$. In particular, we constructed 
$$
\hat{\nu}_l = \int_0^\infty \frac{\lam^l}{l!}\, e^{-\lam}\ \widehat{g}(\lam) d \lam = 
\hat{\pi}_0 \II(l=0) + (1- \hat{\pi}_0) \sum_{k=1}^p \hat{\theta}_k U_k(l),
$$
where  $U_k(l)$ defined in equation \fr{def_Uk},  and then evaluated the weighted $\ell^2$-norm of the difference between the vectors 
$\hat{\bnu} = (\hat{\nu}_0, \hat{\nu}_1, \ldots, )$ and $\bnu = (\nu_0, \nu_1, \ldots)$ of, respectively, the predicted and the observed frequencies
\be \label{Del_nu}
\Delta_{\nu} =  \|\bnu - \hat{\bnu}\|_2^2 /\|\bnu\|^2_2.
\ee
For the estimator proposed in this paper, we tested various computational schemes   
that differ by the strategies for selecting the penalty parameter $\alpha$ in  expression \fr{lasso_for_theta}.
In addition, we compared our estimator with the estimator of Comte and  Genon-Catalot (2015). 
In particular, we considered the following techniques for choosing $\alpha$.

 \begin{itemize}

\item[]   $OPT:$ This estimator is obtained using algorithm presented in Section \ref{sec:implement}.
Parameter $\alpha$   is optimally chosen by minimizing the difference $\|g-\hat{g}\|^2_2$ 
between the true and estimated values of $g$. This estimator represents a benchmark for the proposed procedures
but it is available only in simulation setting but not in practice.

\item[]   $DD_{l2}:$ This estimator is obtained using algorithm presented in Section \ref{sec:implement} 
where parameter $\alpha$ in  \fr{lasso_for_theta}  is chosen by a data driven (DD) criterion. 
The general idea behind such kind of criterion is to measure, as a function of parameter 
$\alpha$, the ability of the estimator to    ``fit'' the observed data, and 
then to choose  $\alpha$  maximizing such kind of  measure. Since we use $\Delta_{\nu}$ 
as a measure of goodness of fit,   estimator $DD_{l2}$ uses the value of $\alpha$ that minimizes
$\Delta_{\nu}$.

\item[]   $DD_{like}:$  This estimator is obtained using algorithm presented in Section \ref{sec:implement} 
where parameter $\alpha$ is derived by maximizing the likelihood   as suggested in \cite{GCV_density_paper}. 
In particular, one can check that the likelihood, given  the data $Y_1,...,Y_n$,  turns out to be 
of the form $\prod_{l=0}^M \hat{\nu}_l^{\nu_l}$,  where $M= \max_i   Y_i$.

\item[]   $NDE:$ This is the Nonparametric Density Estimator presented in Comte and  Genon-Catalot (2015).  
The authors kindly provided the code.
\end{itemize}

\noindent
The set of test functions represents different situations inspired by the real 
data problem described in the next  Section. In particular, we consider the following nine test functions: 
\begin{enumerate}
\item  a gamma density $g(\lambda)=\Gamma(\lambda; 3,1)$
\item a mixed gamma density $g(\lambda)=0.3\Gamma(\lambda; 3,0.25)+0.7\Gamma(\lambda; 10,0.6)$
\item an exponential density $g(\lambda)=\Gamma(\lambda; 1,2)$
\item a Weibull density $g(\lambda)=\theta p^{-\theta} x^{\theta-1} \exp{-(x/\theta)^\theta} 1_{x>0}$, with $p=3$ and $\theta=2$
\item a Gaussian density $g(\lambda)=N(\lambda; 80,1)$
\item a mixed gamma density $g(\lambda)=0.3\Gamma(\lambda; 2,0.3)+0.7\Gamma(\lambda; 40,1)$
\item a delta contaminated gamma density $g(\lambda)=0.3\delta(\lambda)+0.7\Gamma(\lambda; 40,1)$
\item a delta contaminated Gaussian density $g(\lambda)=0.2\delta(\lambda)+0.8 N(\lambda; 80,8^2)$
\item a delta contaminated Gaussian density $g(\lambda)=0.2\delta(\lambda)+0.8 N(\lambda; 20,4^2)$ 
\end{enumerate}

\begin{figure}[htp]
  \centering
  \caption{The true density (red) and $DD_{like}$ estimators (blue) 
obtained in the first 10 simulation runs with  sample size $n=5000$}
  \begin{tabular}{ccc}
    \includegraphics[width=50mm]{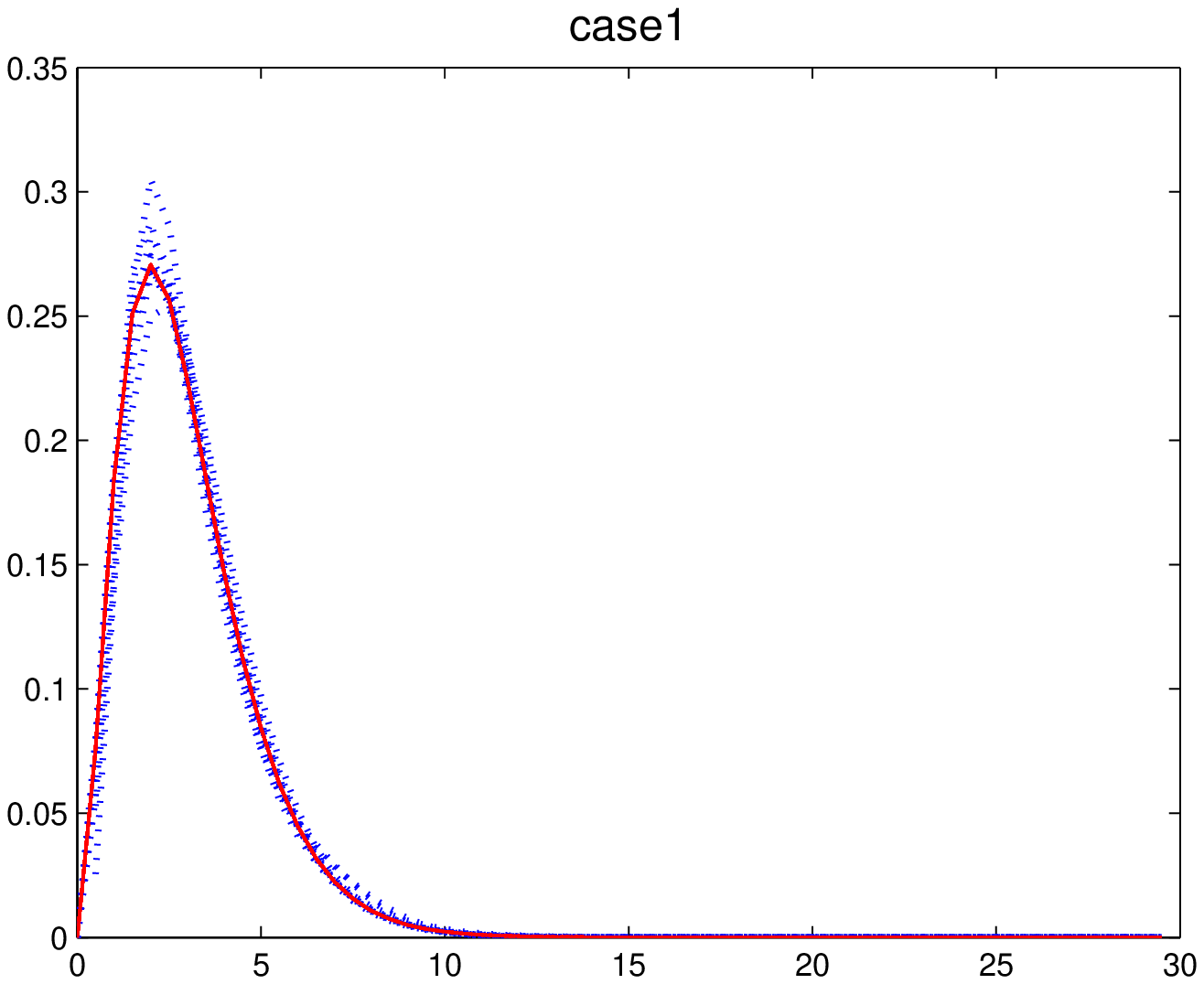}&
    \includegraphics[width=50mm]{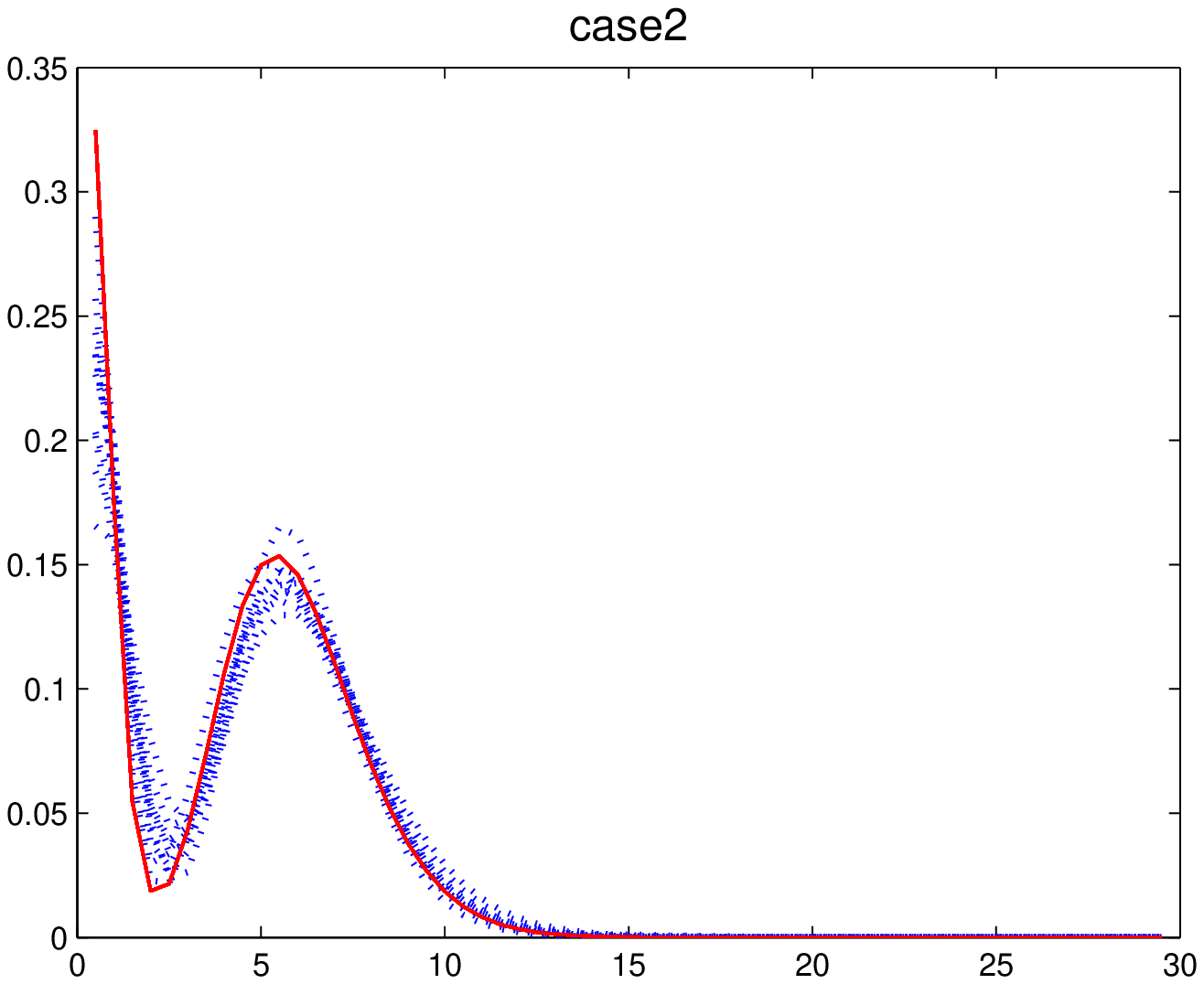}&
    \includegraphics[width=50mm]{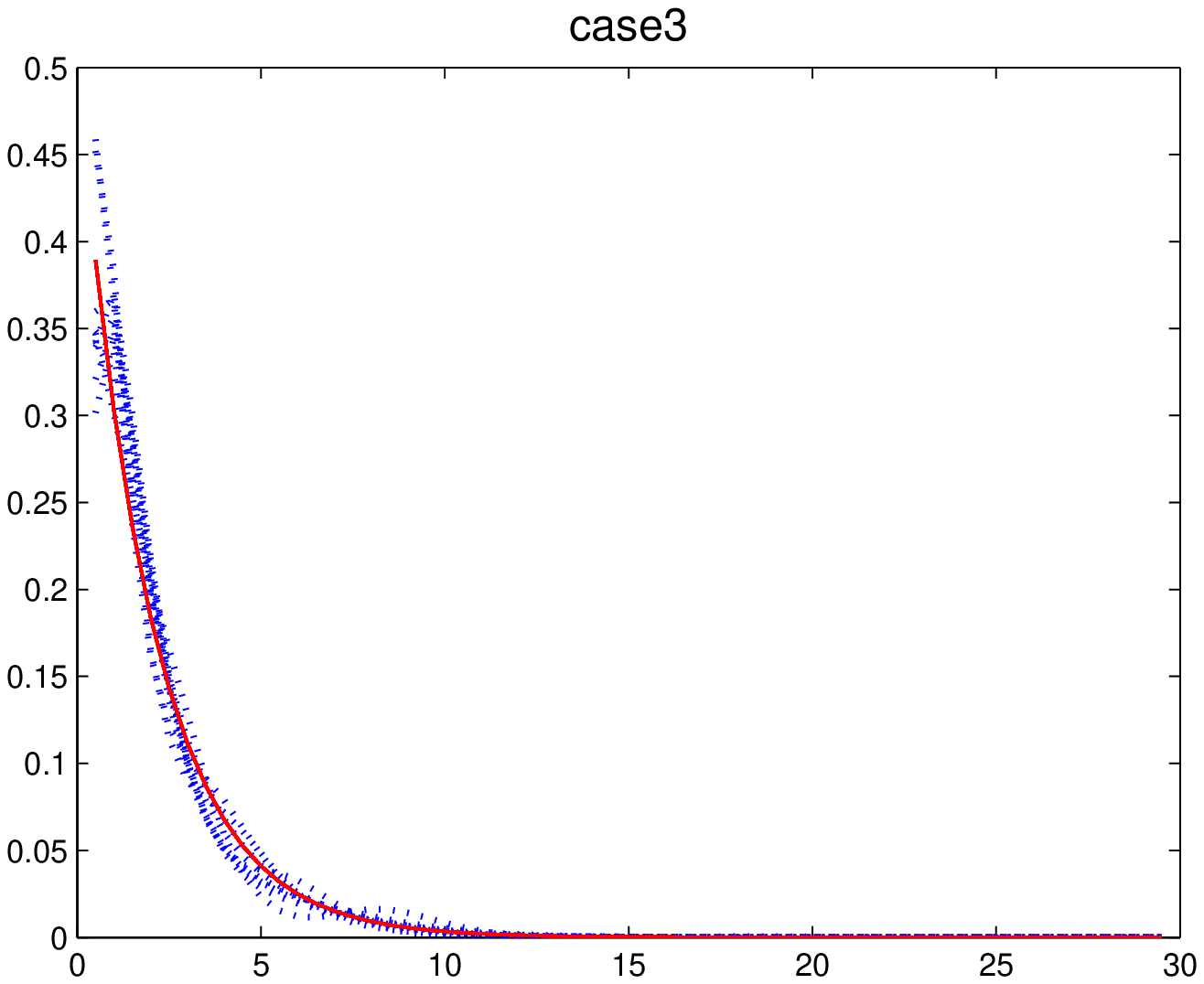}\\
     \includegraphics[width=50mm]{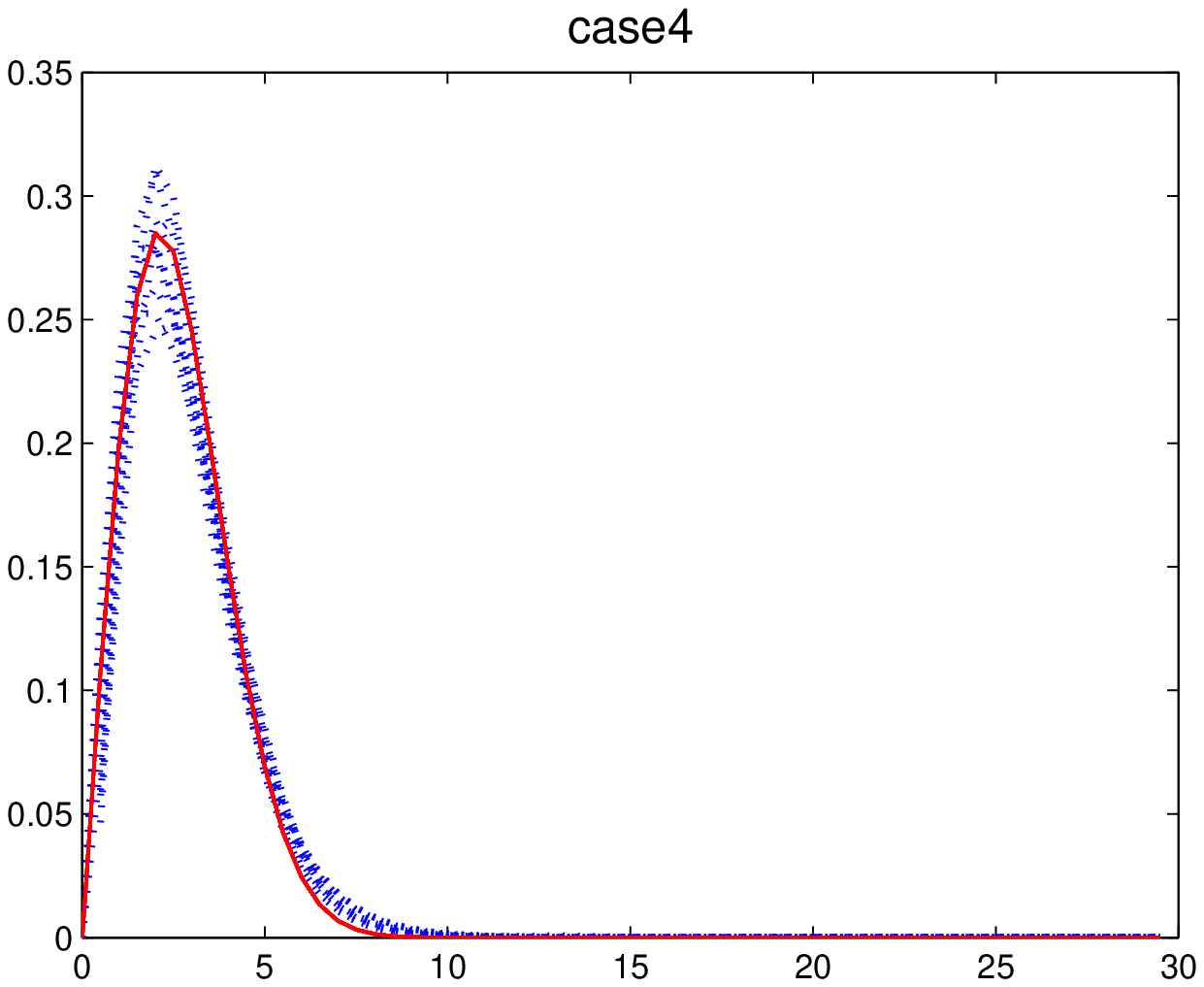}&
    \includegraphics[width=50mm]{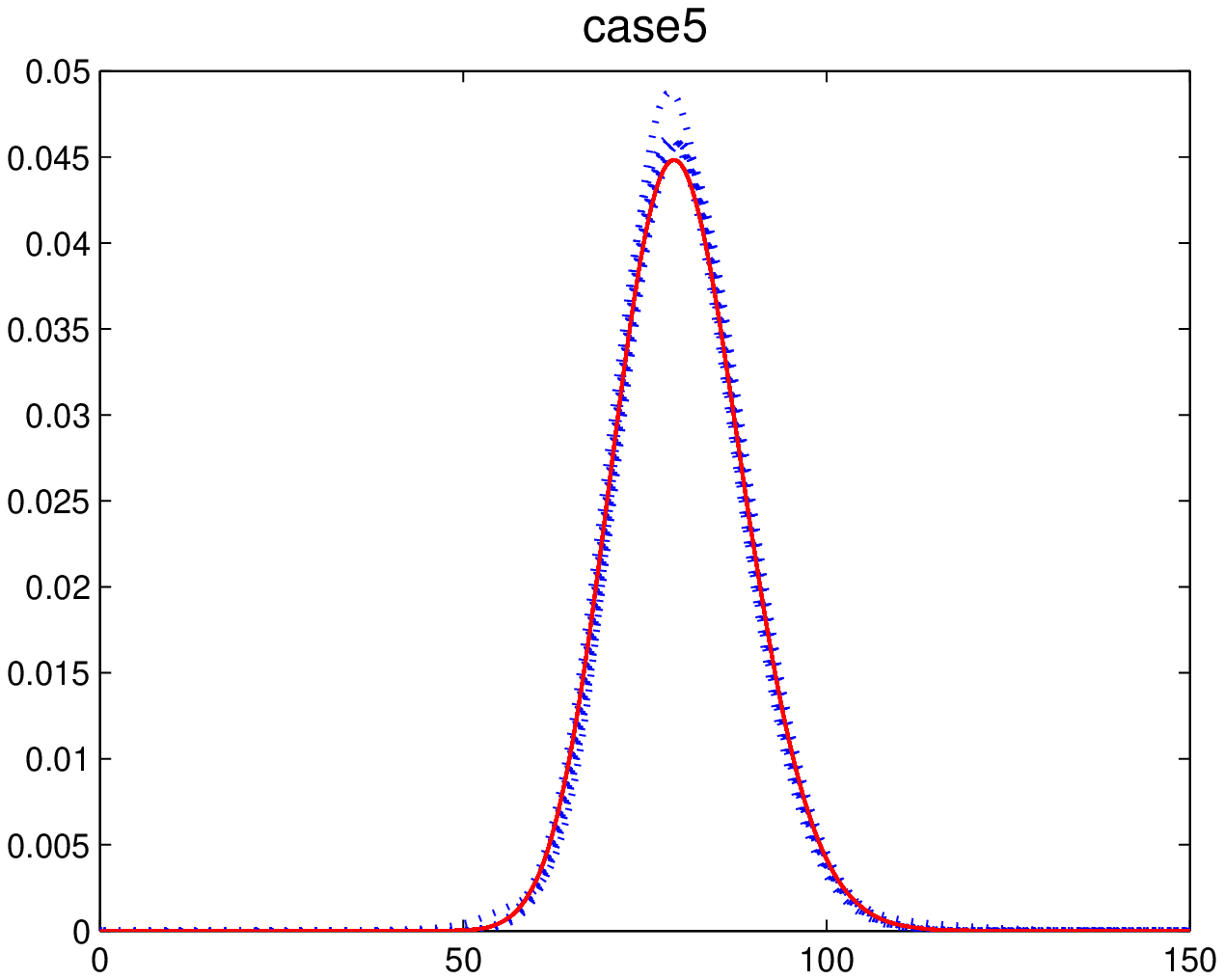}&
    \includegraphics[width=50mm]{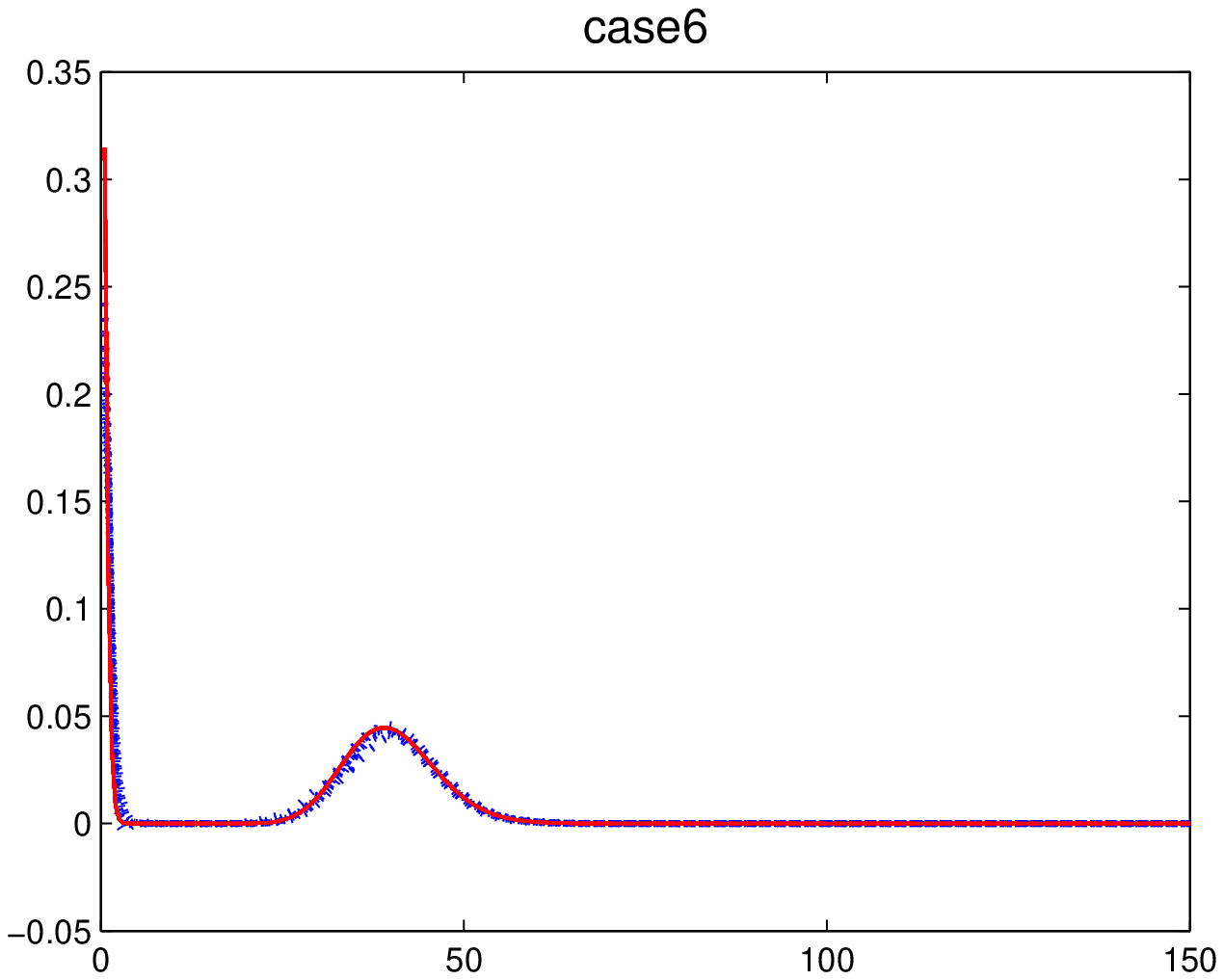}\\
     \includegraphics[width=50mm]{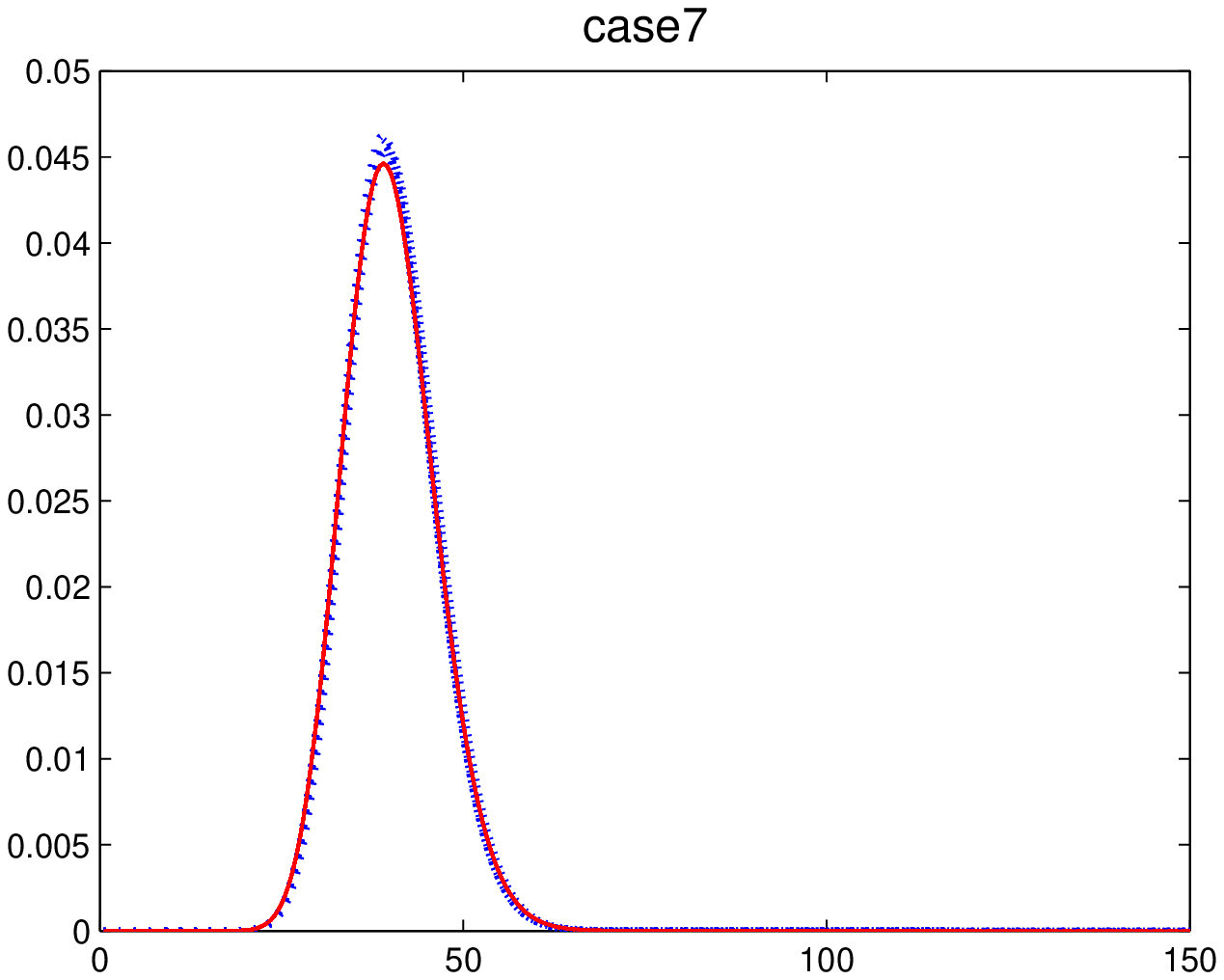}&
    \includegraphics[width=50mm]{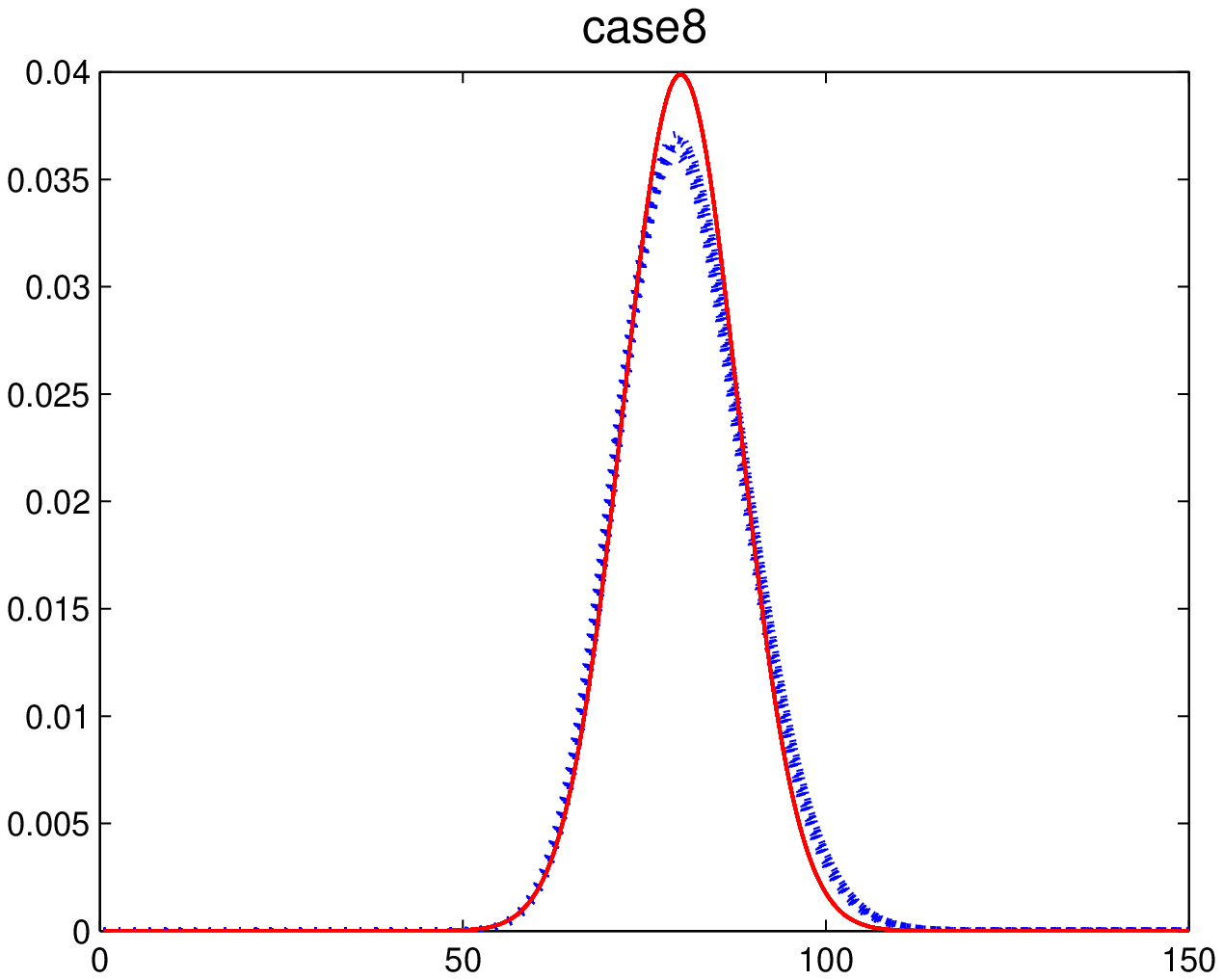}&
    \includegraphics[width=50mm]{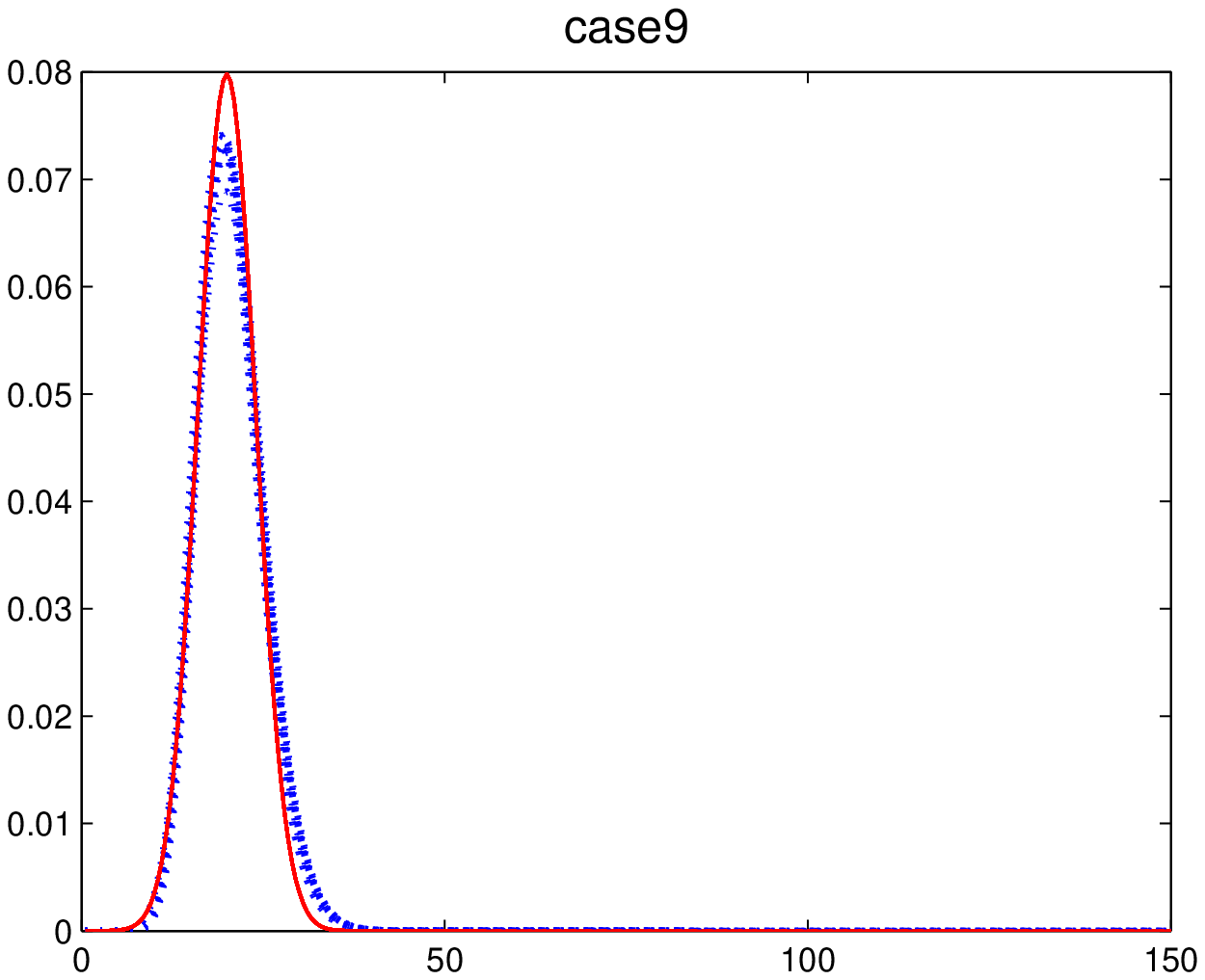}\\
  \end{tabular}\label{figure1}
\end{figure}

The first four test functions have been analyzed in Comte and  Genon-Catalot (2015) and represent cases 
where most of the data is concentrated near zero. The  5-th test function corresponds to the situation where 
 most of the data is concentrated away from zero. The   last four test functions represent the  mixtures  of 
the two previous scenarios. All nine densities are showed in Figure  \ref{figure1}.

Tables 1, 3 and 5 below display the average values of  $\Delta_g$ defined in \fr{Del_g} 
while Tables 2, 4 and 6 report $\Delta_\nu$ given by \fr{Del_nu} (with the standard deviations in parentheses) 
over 100 different realizations of data  $Y_i \sim Poisson(\lambda_i)$, $i=1,..,n$, where $n=10000$ for Tables 1 and 2, 
$n=5000$ for Tables 3 and 4 and $n=1000$ for Tables 5 and 6. The dictionary was constructed as a collection of the gamma pdfs  
\fr{def_phi} where parameters $(a_k, b_k)$  belong to the Cartesian product of vectors $a=[2,3,4,\cdots,150]$ 
and $b=[0.1,0.15,\cdots, 0.9,0.95]$, so that $\phi_k(0)=0$. We chose the grid step   $h=0.5$.

\begin{figure}[htp]
  \centering
  \caption{Sample frequencies (red) and estimated frequencies (blue) 
obtained in the first 10 simulation runs with  sample size $n=5000$}
  \begin{tabular}{ccc}
    \includegraphics[width=50mm]{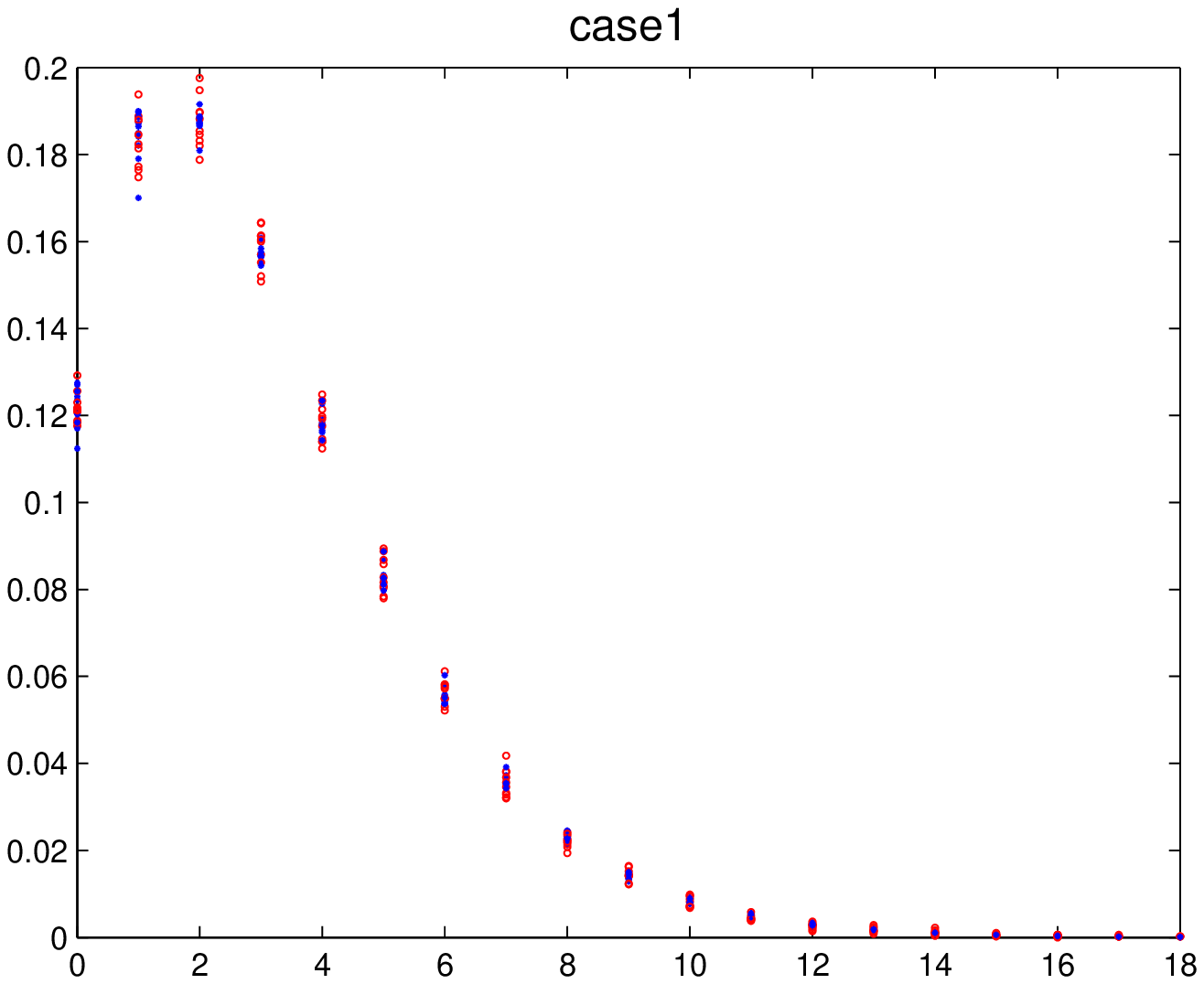}&
    \includegraphics[width=50mm]{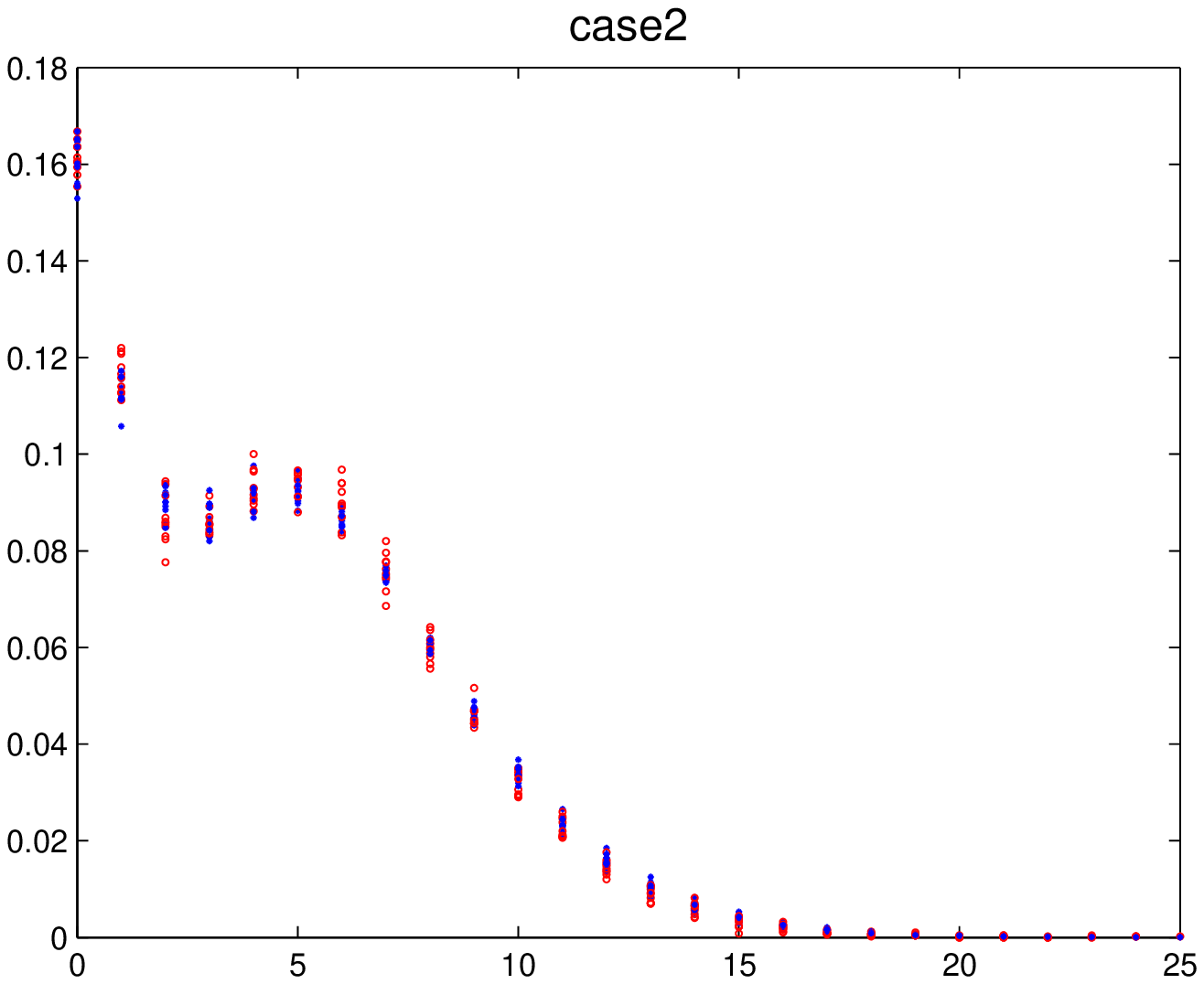}&
    \includegraphics[width=50mm]{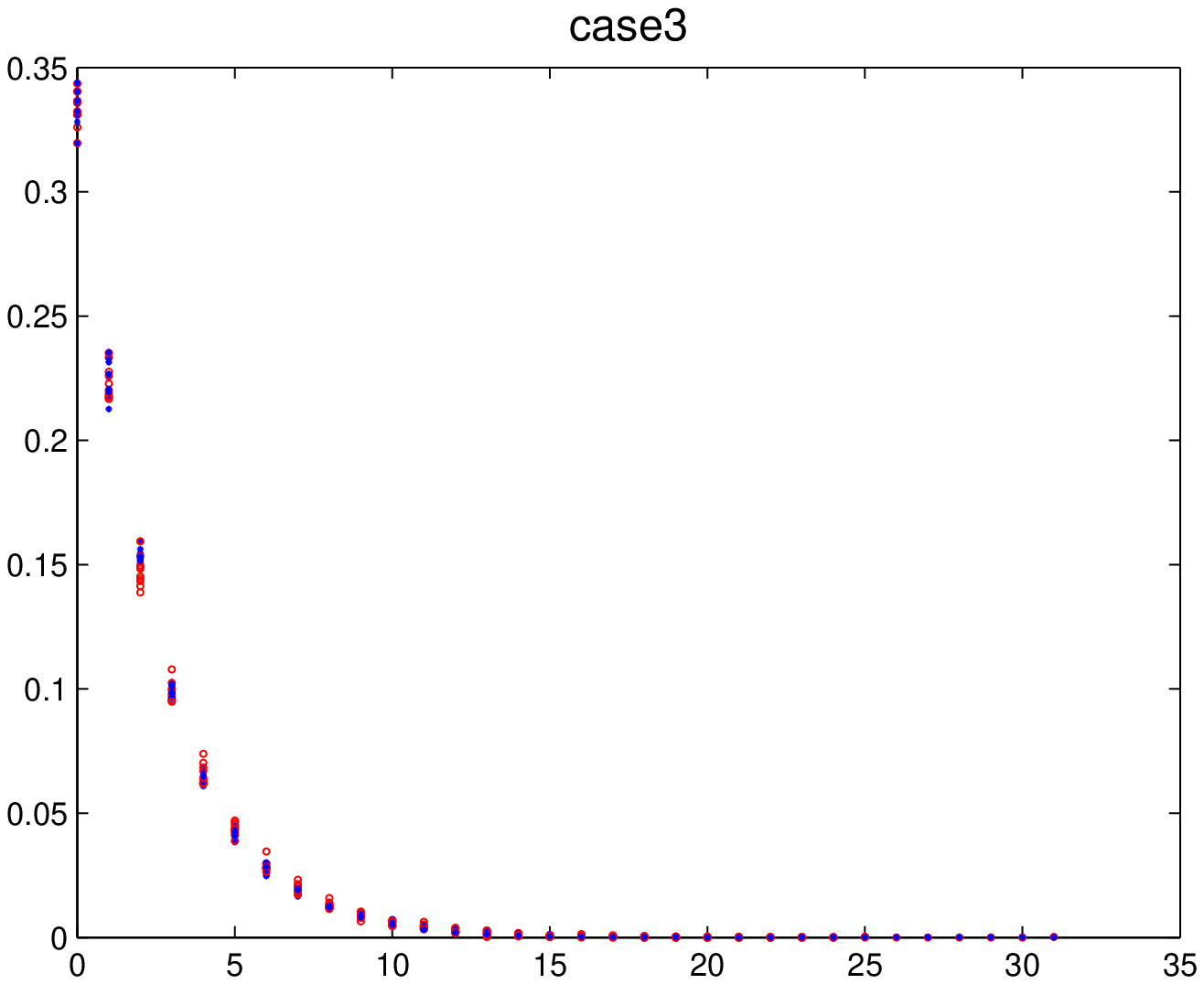}\\
     \includegraphics[width=50mm]{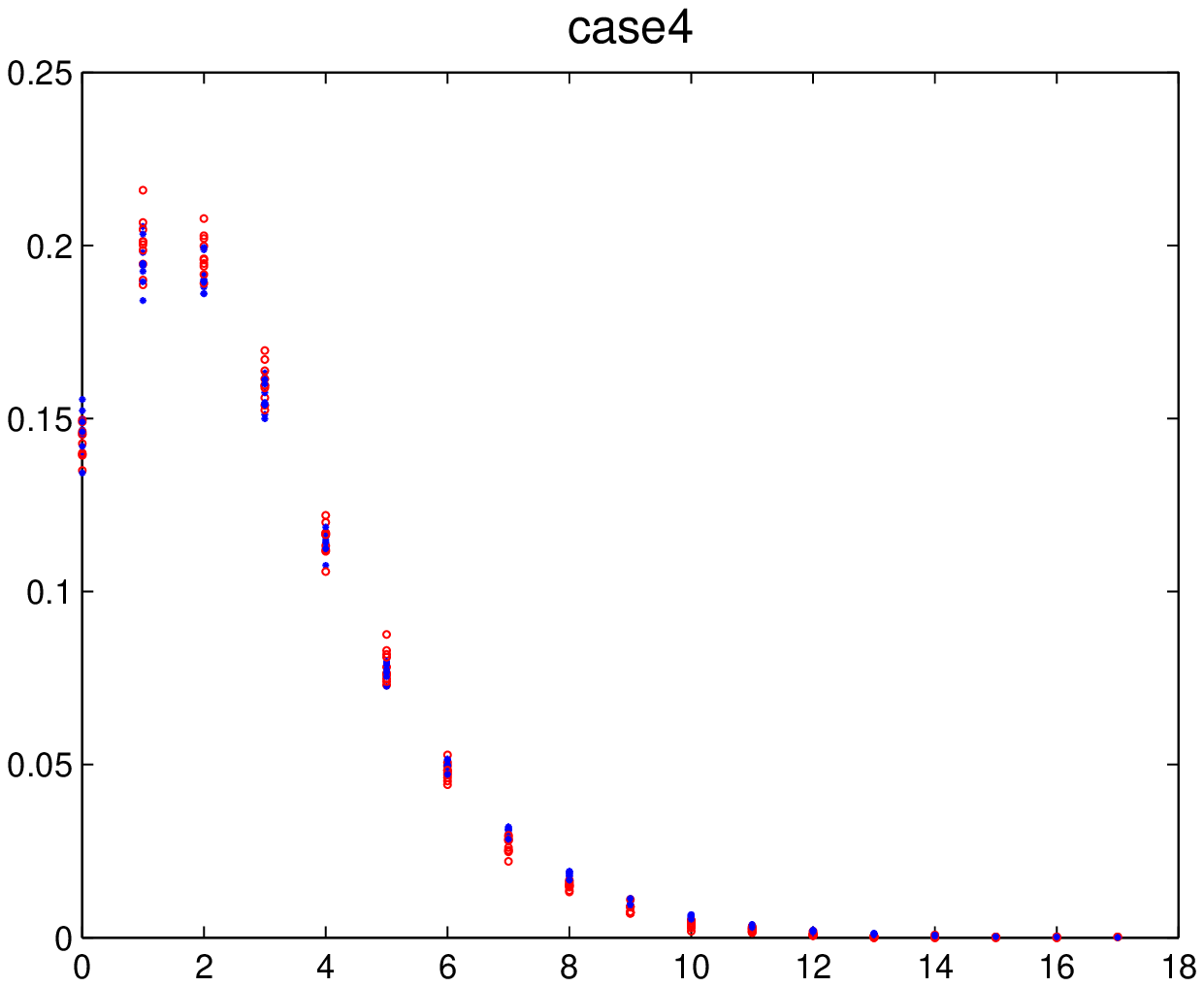}&
    \includegraphics[width=50mm]{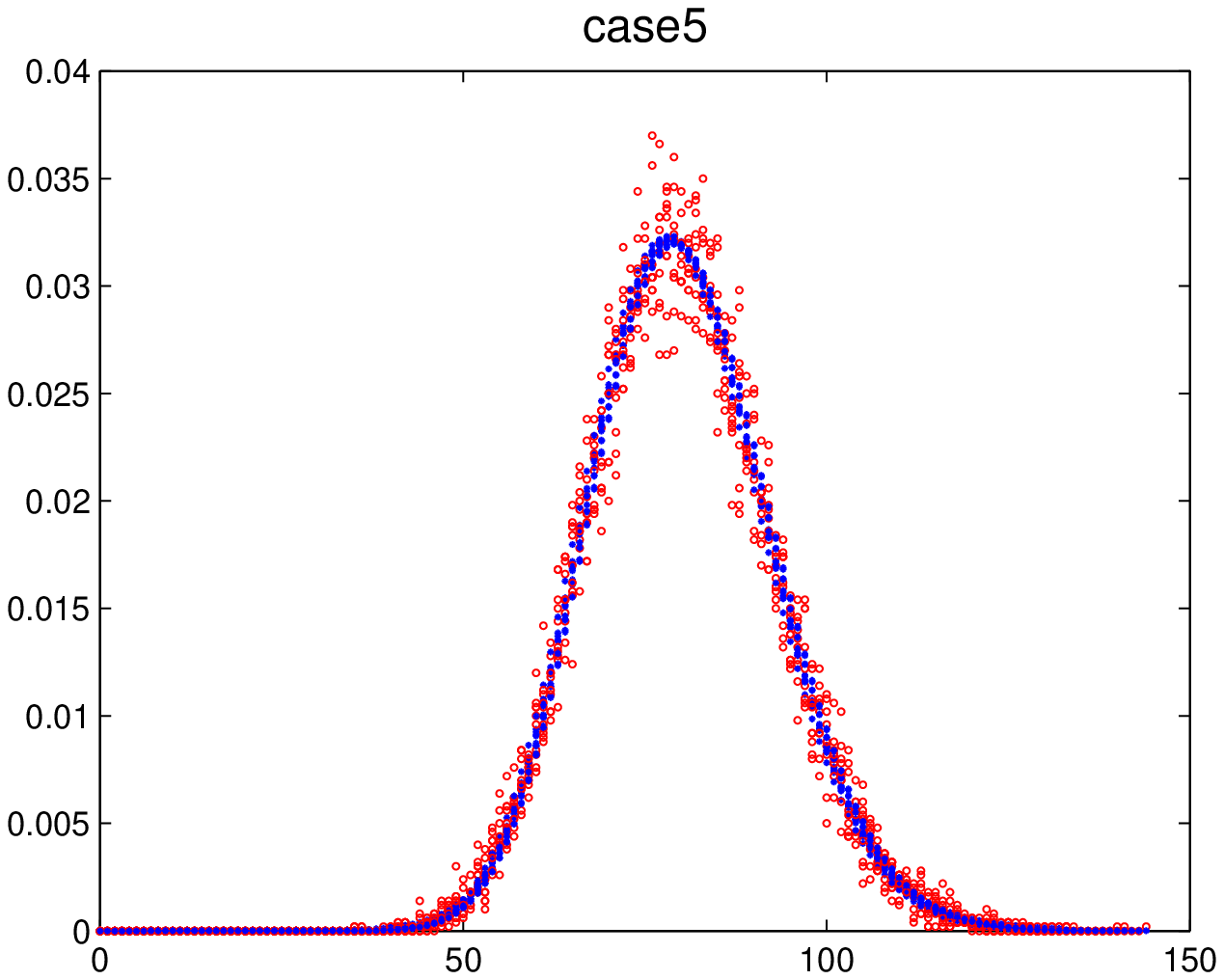}&
    \includegraphics[width=50mm]{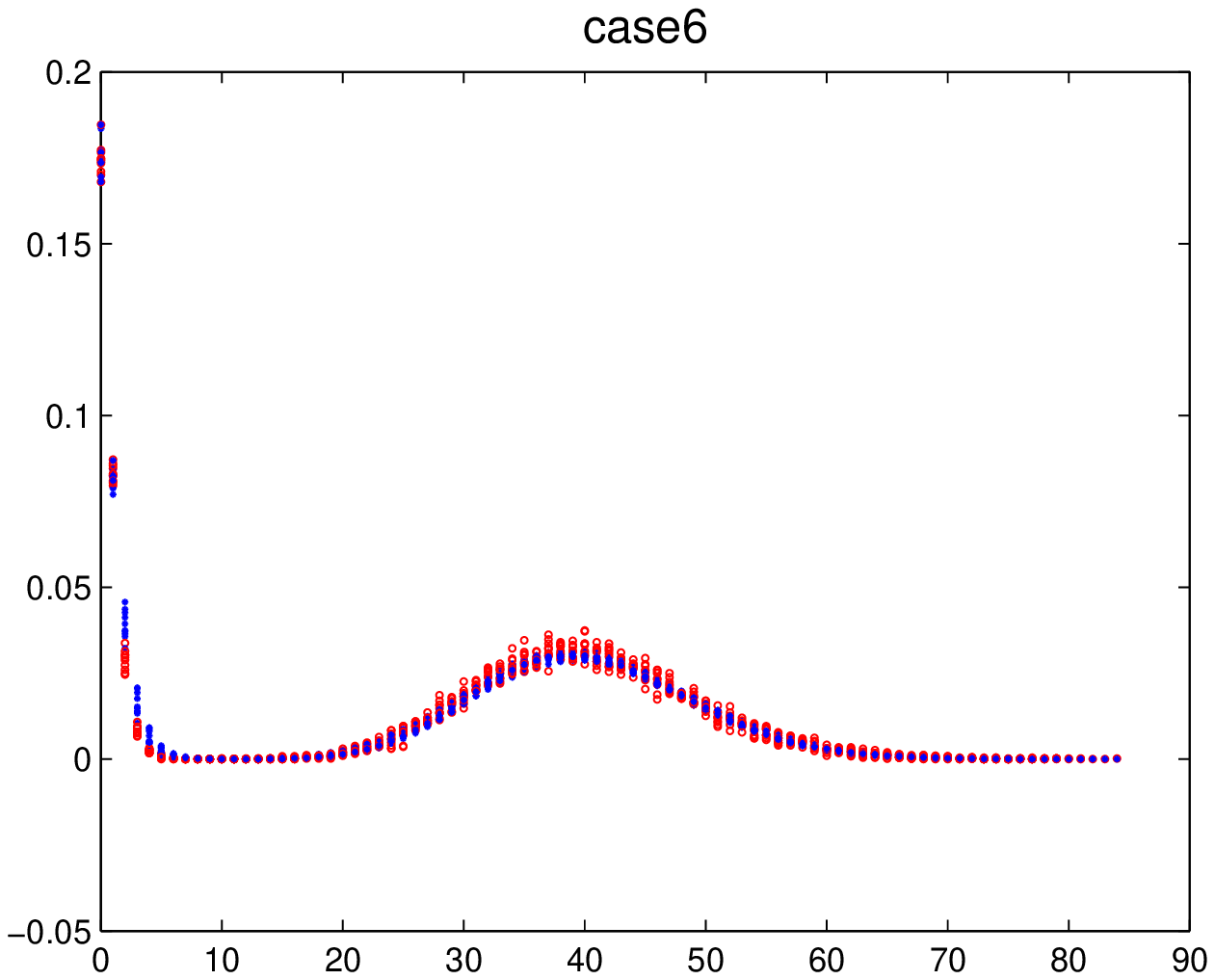}\\
     \includegraphics[width=50mm]{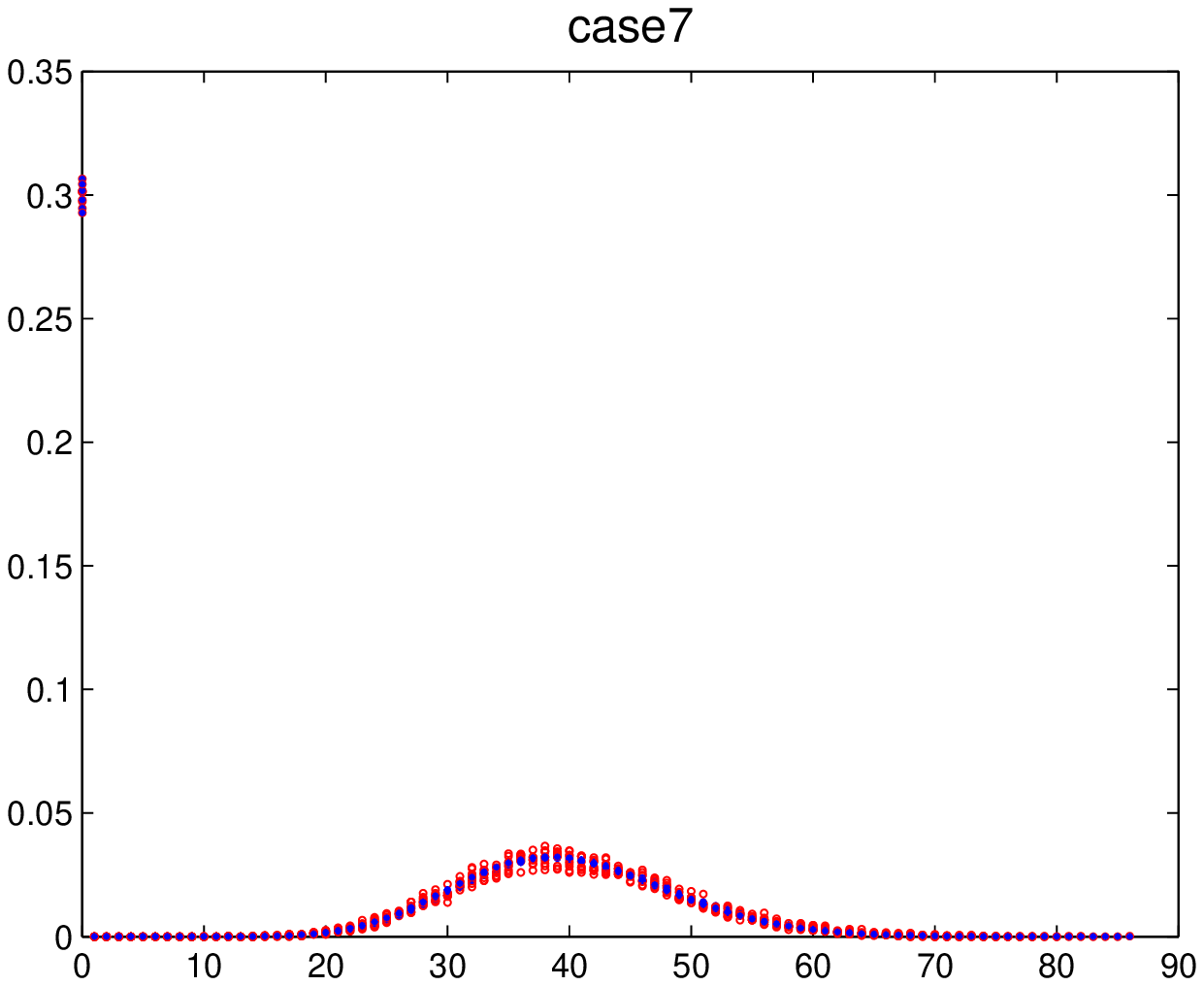}&
    \includegraphics[width=50mm]{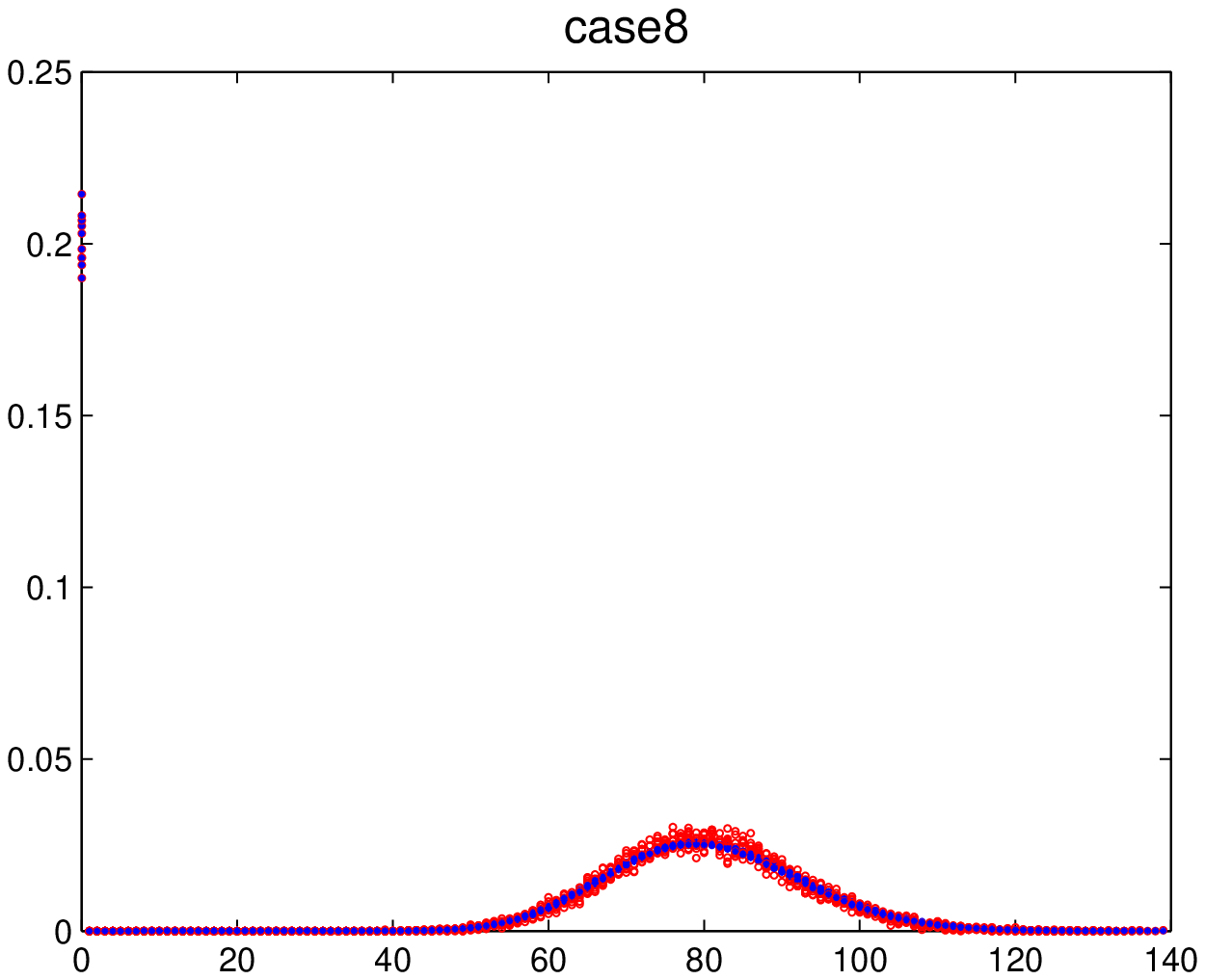}&
    \includegraphics[width=50mm]{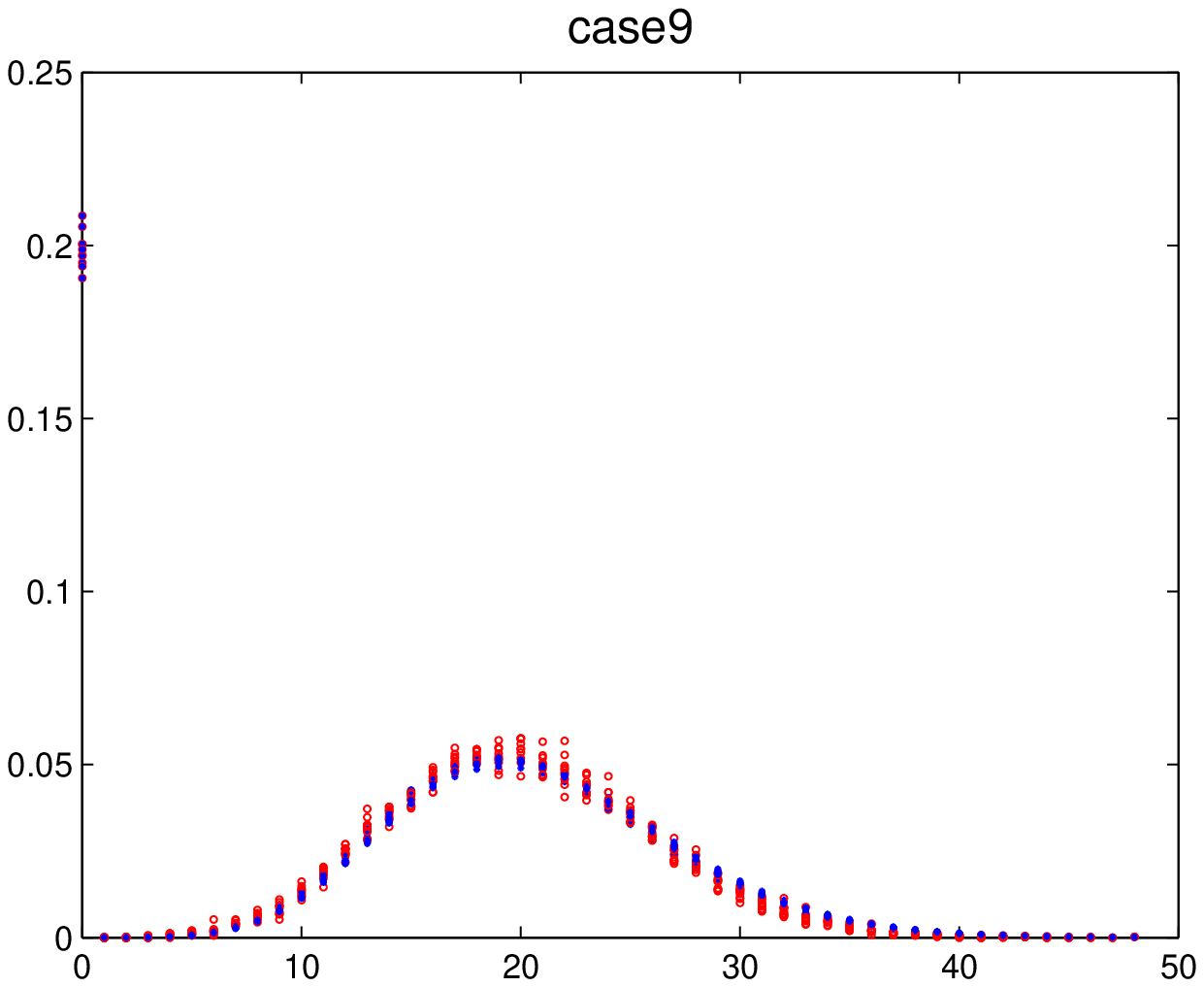}\\
  \end{tabular}\label{figure2}
\end{figure}

 As it is expected, performances of all estimators deteriorate when   $n$ decreases, although not very significantly.
For a fixed sample size, estimator $OPT$ is the most precise in terms of  $\Delta_g$ as a direct consequence of its definition,
however, estimator $DD_{like}$ is always comparable. Estimator  $DD_{l2}$ has similar performance to   $DD_{like}$ 
except for cases 2 and 6 where the underlying densities are bimodal and, hence,  data can be explained by a variety of 
density mixtures. In conclusion, apart from $OPT$ which is not available in the case of real data, estimator $DD_{like}$  
turns out to be the most accurate in terms of both  $\Delta_g$ and  $\Delta_\nu$. For completeness, 
Figures \ref{figure1} and \ref{figure2} exhibit some reconstructions obtained using estimator $DD_{like}$ in case of $n=5000$.

Finally, we should mention that $NDE$ is a projection estimator that uses only the first few Laguerre functions. For this reason, it fails  
to adequately represent a density function that corresponds to the situation where values $\lam_i$, $i=1, \cdots, n$, are 
concentrated away from zero, as it happens in case 5 (where $NDE$ returns zero as an estimator) and case 6 
(where $NDE$ succeeds in reconstructing only the first part of the density near zero). Also, note that $NDE$ errors are not displayed 
 for  cases 7, 8 and 9   since  this estimator   is not defined for delta contaminated densities.

\section {Application to evaluation of the density of the Saturn ring}
\label{sec:real data}
\setcounter{equation}{0}

\begin{figure}[htp]
  \centering
 \caption{The first panel: Names of Saturn's rings, courtesy  {\tt science.nasa.gov}.  
The second panel: the means of the binned   total data set (100 observations per bin). 
The third panel:  the means  (red) and the variances (blue)  of the binned data}
  \begin{tabular}{c}
    \includegraphics[width=150mm,height=65mm]{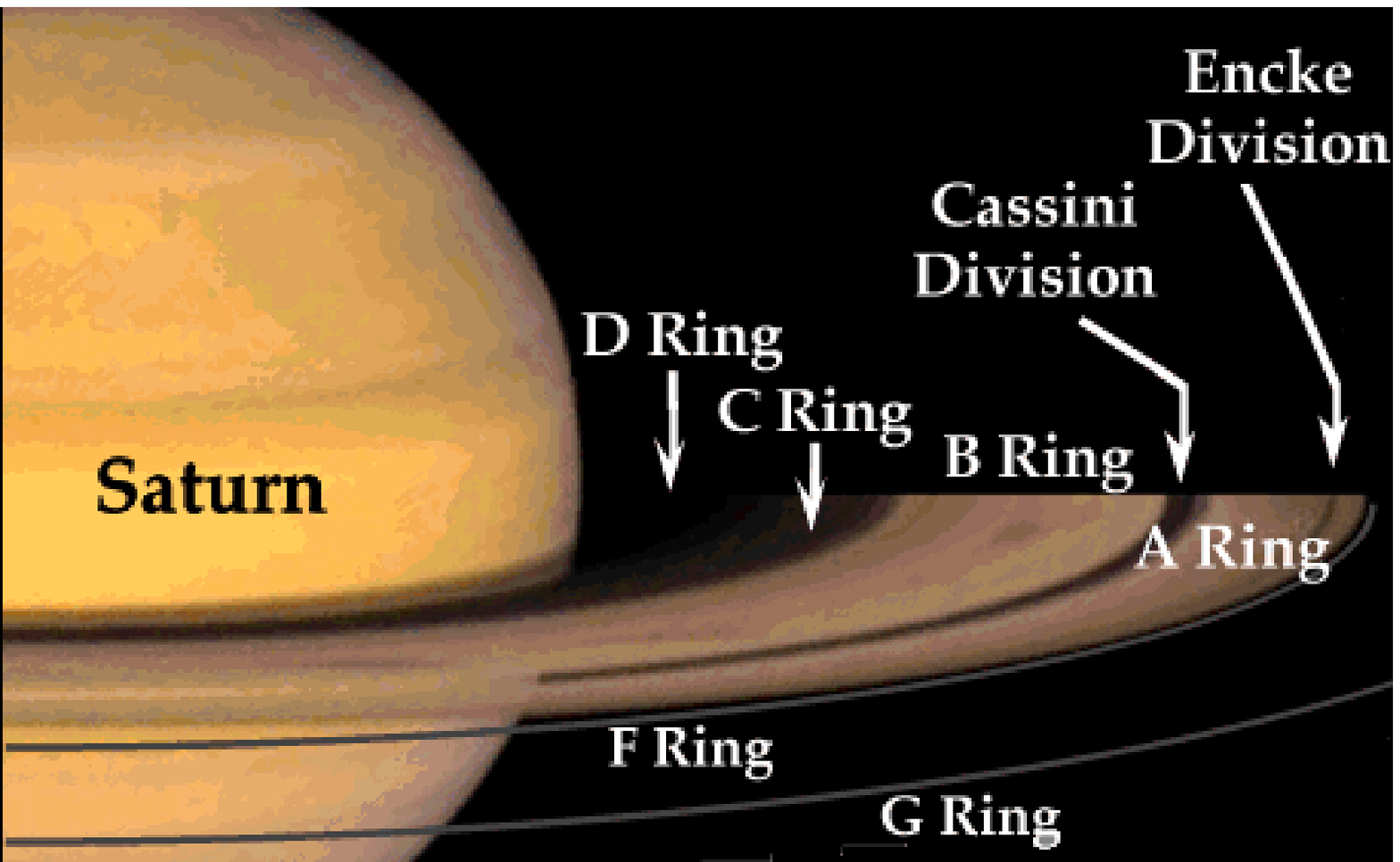}  \label{fig_real_data} \\
     \includegraphics[width=150mm,height=65mm]{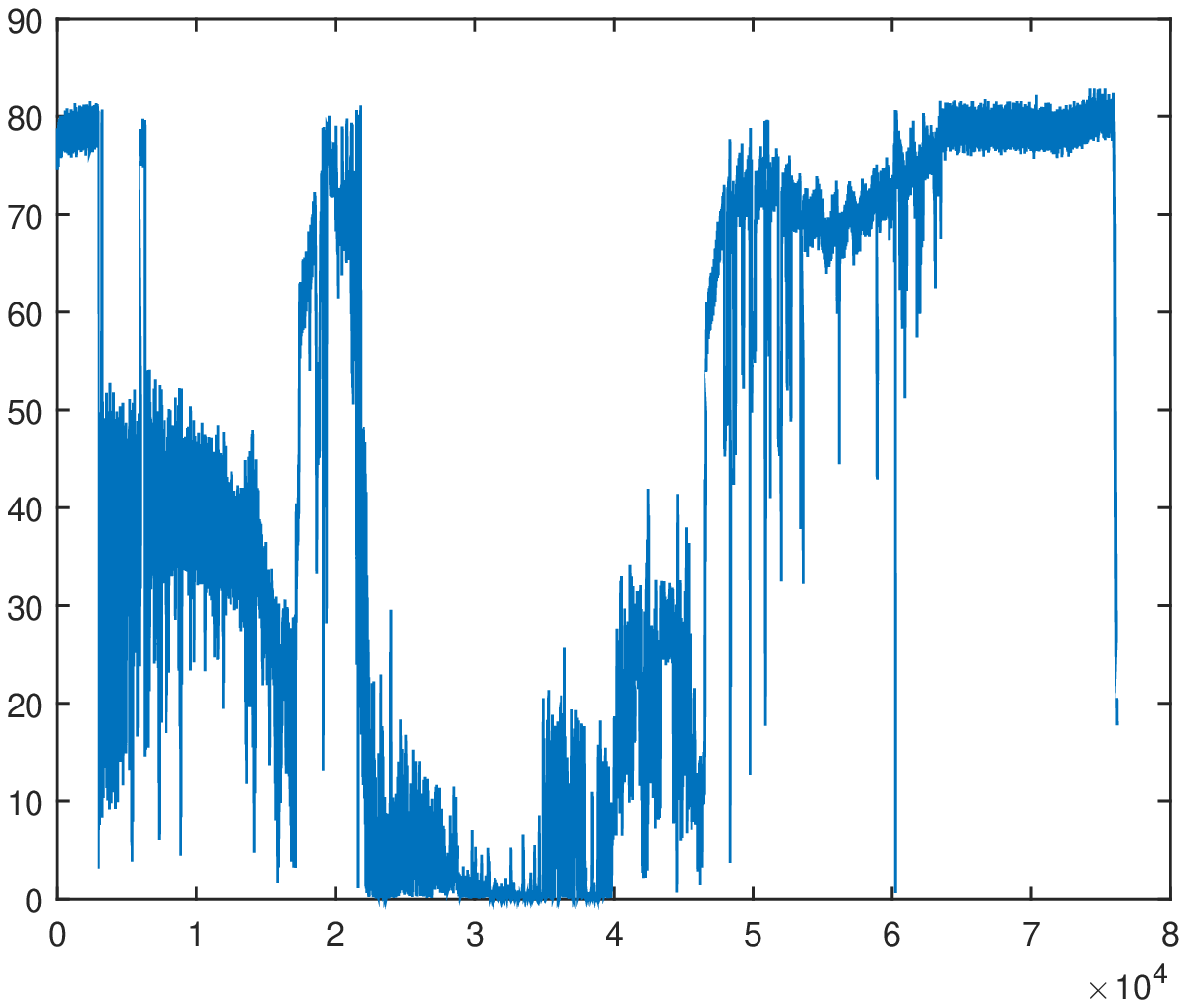}\\
    \includegraphics[width=150mm,height=65mm]{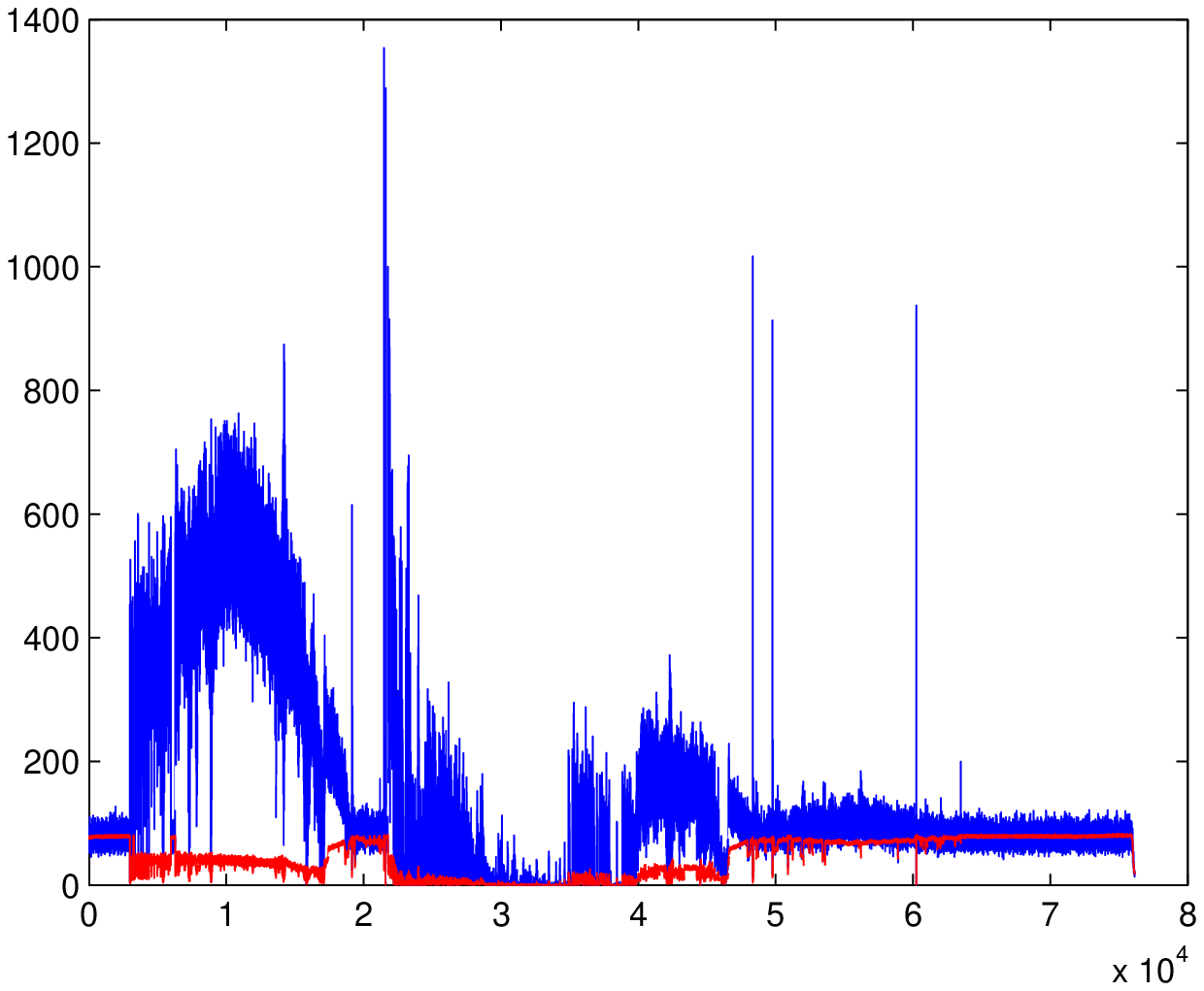} \\
    \end{tabular}
\end{figure}

The Saturn's rings system can be broadly grouped into two categories: dense rings (A, B, C) and tenuous rings (D, E, G)
(see the first panel of Figure~\ref{fig_real_data}). 
The Cassini Division is a ring region  that separates the A and B rings.
The study of structure within Saturn's rings originated with   Campani, who observed in 1664 
that the inner half of the disk was brighter than the outer half. Furthermore,
in 1859,    Maxwell proved that the rings could not be solid or liquid but were 
instead made up of an indefinite number of   particles of various sizes, each on its own orbit about Saturn.
Detailed ring structure was revealed for the first time, however, by the 1979 Pioneer and 
1980-1981 Voyager encounters with Saturn.   Images were taken at close range, by stellar 
occultation (observing the flickering of a star as it passes behind the rings), 
and by radio occultation (measuring the attenuation of the spacecraft's radio signal 
as it passes behind the rings as seen from Earth) (see, e.g., Esposito et al, 2004).
By analyzing the intensity of star light while it is passing through Saturn's rings, 
astronomers can gain insight into properties telescopes cannot visually determine.  
Each  sub-region in the rings  has its own  associated distinct distribution of 
of the density and sizes   of the particles constituting the sub-region. This distribution   
uniquely determines the amount of light which is able to pass from a star (behind the rings) 
to the photometer.

Our data  $Y_i,\ i=1, \cdots, n,$ come  from sets of observations   of stellar occultations  
recorded by the Cassini UVIS high speed photometer and contains $n = 7\,615\,754$ photon counts
at different radial points, located   at  0.01-0.1 kilometer increments, on the Saturn's rings plane
(see the second panel of Figure~\ref{fig_real_data}).
With no outside influences, these photon counts should follow  the Poisson distribution, however, 
obstructions imposed by the particles in the rings cause their distribution to deviate from Poisson.
Indeed, if data were Poisson distributed, then its mean would be approximately equal to its variance for every sub-region.
However,  as the third panel of  Figure~\ref{fig_real_data} 
shows, observations $Y_i$ have significantly higher variances than means. 
The latter is due to the fact that,  although  for each $i=1, \cdots, n$,
the photon counts $Y_i$ are $Poisson (\lam_i)$, the values of $\lam_i$, $i=1, \cdots, n$, are
extremely  varied and, specifically, cannot be modeled as values of a continuous function. 
In fact, intensities  $\lam_i$, $i=1, \cdots, n$, are best described as random variables
with an unknown underlying pdf $g(\lam)$.
In addition, if the ring region contains a significant proportion of large  particle,
those particles can completely block of the light leading to zero photon counts.
For this reason, we allow  $g(\lam)$ to possibly contain a  non-zero mass at $\lam=0$,
hence, being of the form \fr{pdf_g}. The shape of $g(\lam)$ allows one to determine the density and 
distribution of the sizes of the particles of a respective sub-region of  the Saturn rings.
This information, in turn, should shed light on the question of the origin of the rings 
as well as how they reached their current configuration.


%

\begin{figure}[htp]    
  \centering
 \caption{Left panels: segments  of data. Right panels: sample frequencies (blue) and  
estimated frequencies with the penalty parameter  obtained by $DD_{like}$ criterion (red).
$\Delta_\nu = 0.0128$ (top panel),  $\Delta_\nu = 0.0159$ (middle panel),  $\Delta_\nu = 0.0022$,
 $\hat{\pi}_0=0$ for all three cases.
(bottom panel)}
  \begin{tabular}{c}
    \includegraphics[width=150mm,height=50mm]{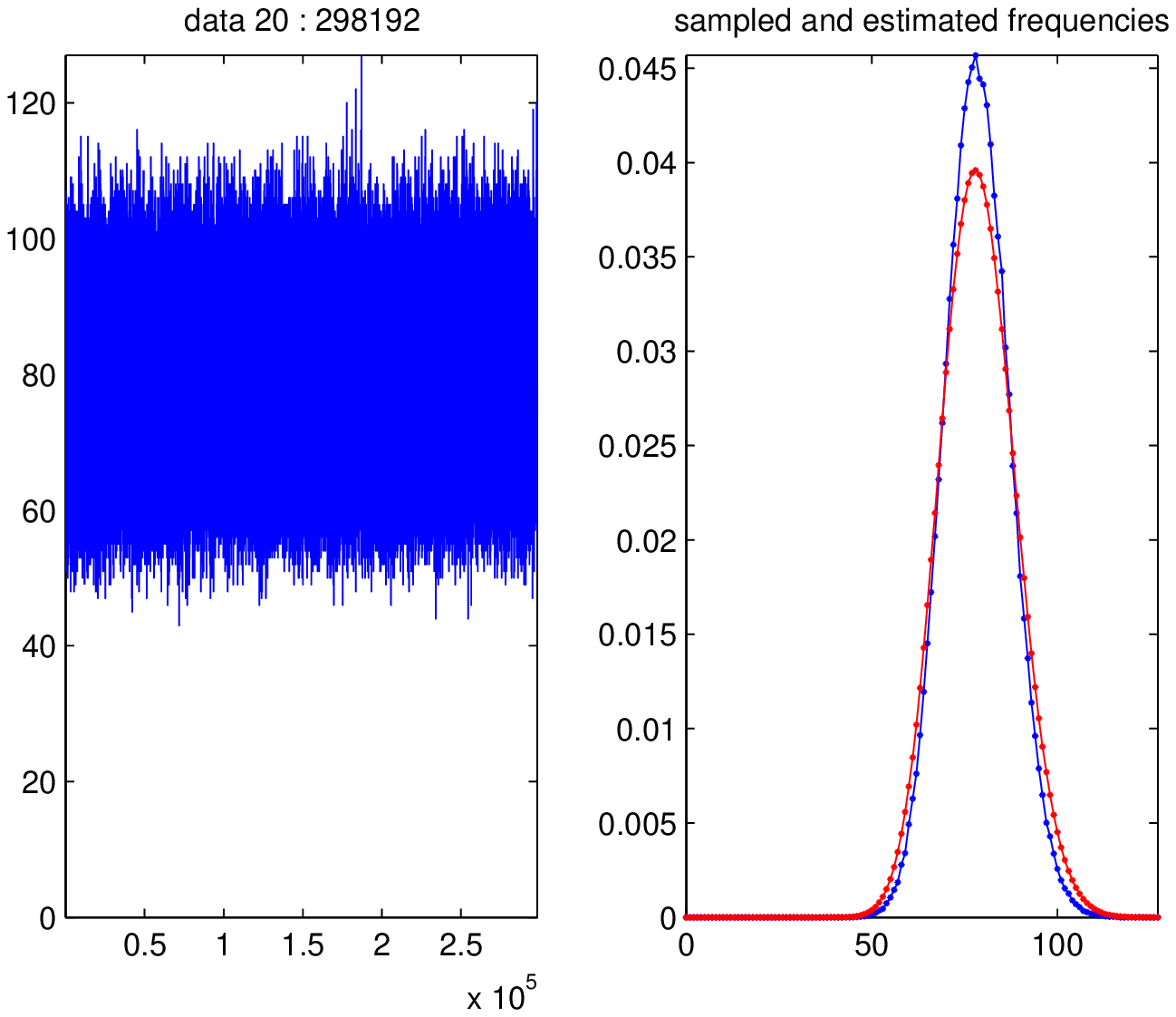}  \label{fig_real_data_pieces1} \\
     \includegraphics[width=150mm,height=50mm]{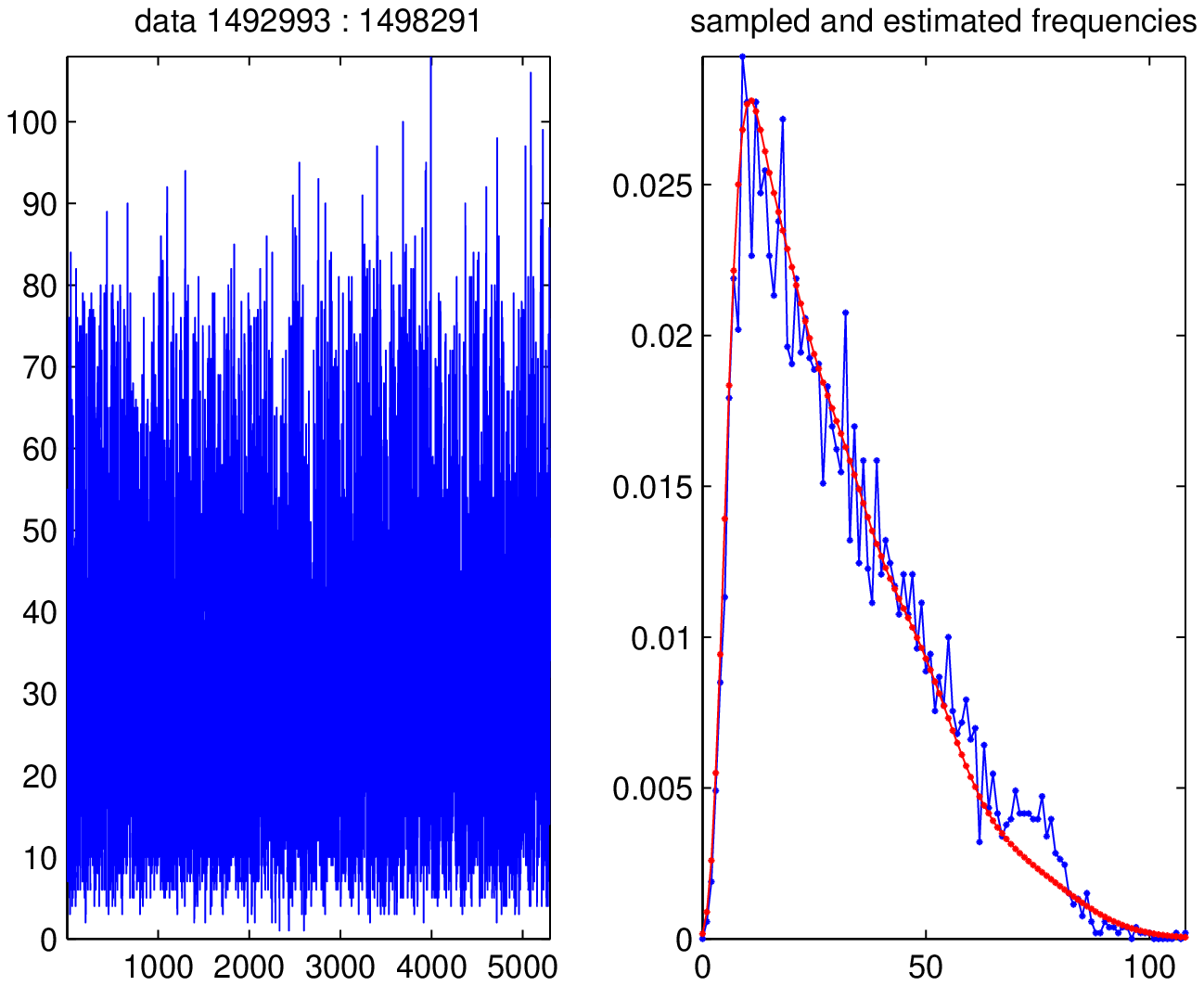}\\
      \includegraphics[width=150mm,height=50mm]{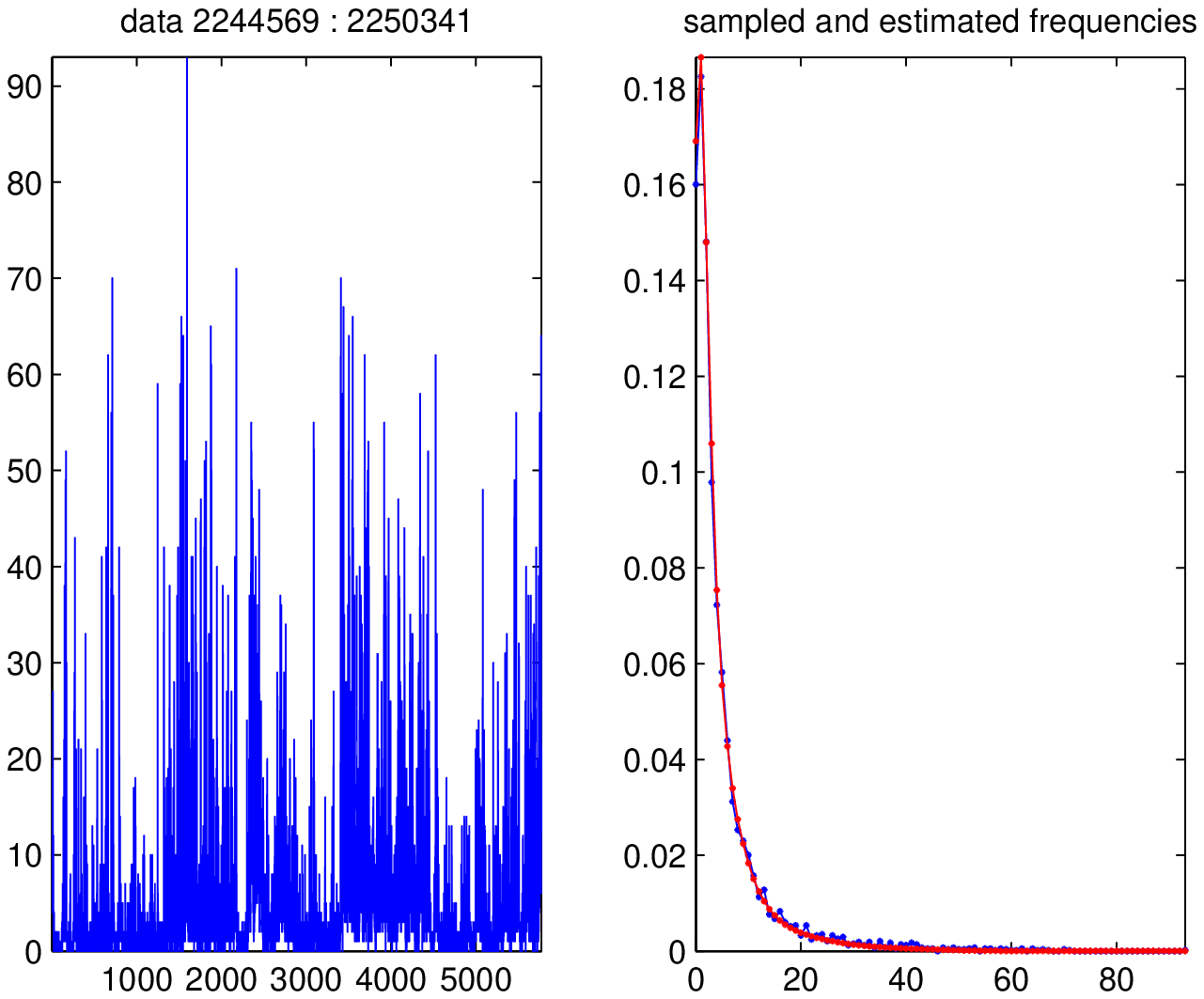}
  \end{tabular}
\end{figure}

\begin{figure}[htp]    
  \centering
 \caption{Left panels: segments  of data. Right panels: sample frequencies (blue) and  
estimated frequencies with the penalty parameter  obtained by $DD_{like}$ criterion (red).
$\Delta_\nu = 0.0229$  and  $\hat{\pi}_0=0.5059$ (top panel),  $\Delta_\nu = 0.003$  and  $\hat{\pi}_0=0.2463$ (middle panel),  
$\Delta_\nu = 0.0095$  and  $\hat{\pi}_0=0$ (bottom panel)}
  \begin{tabular}{c}      
     \includegraphics[width=150mm,height=50mm]{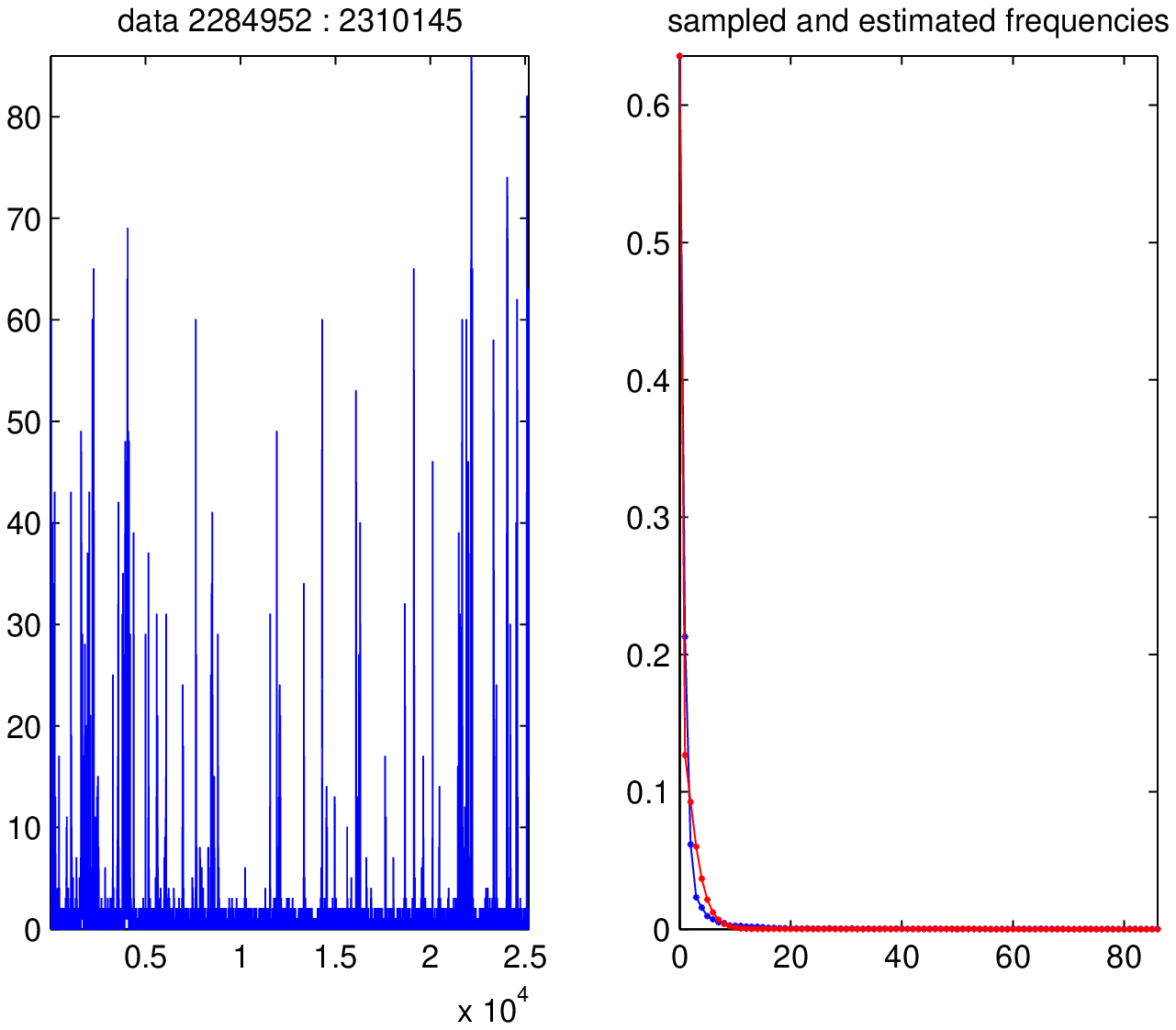}\\ \label{fig_real_data_pieces2}
         \includegraphics[width=150mm,height=50mm]{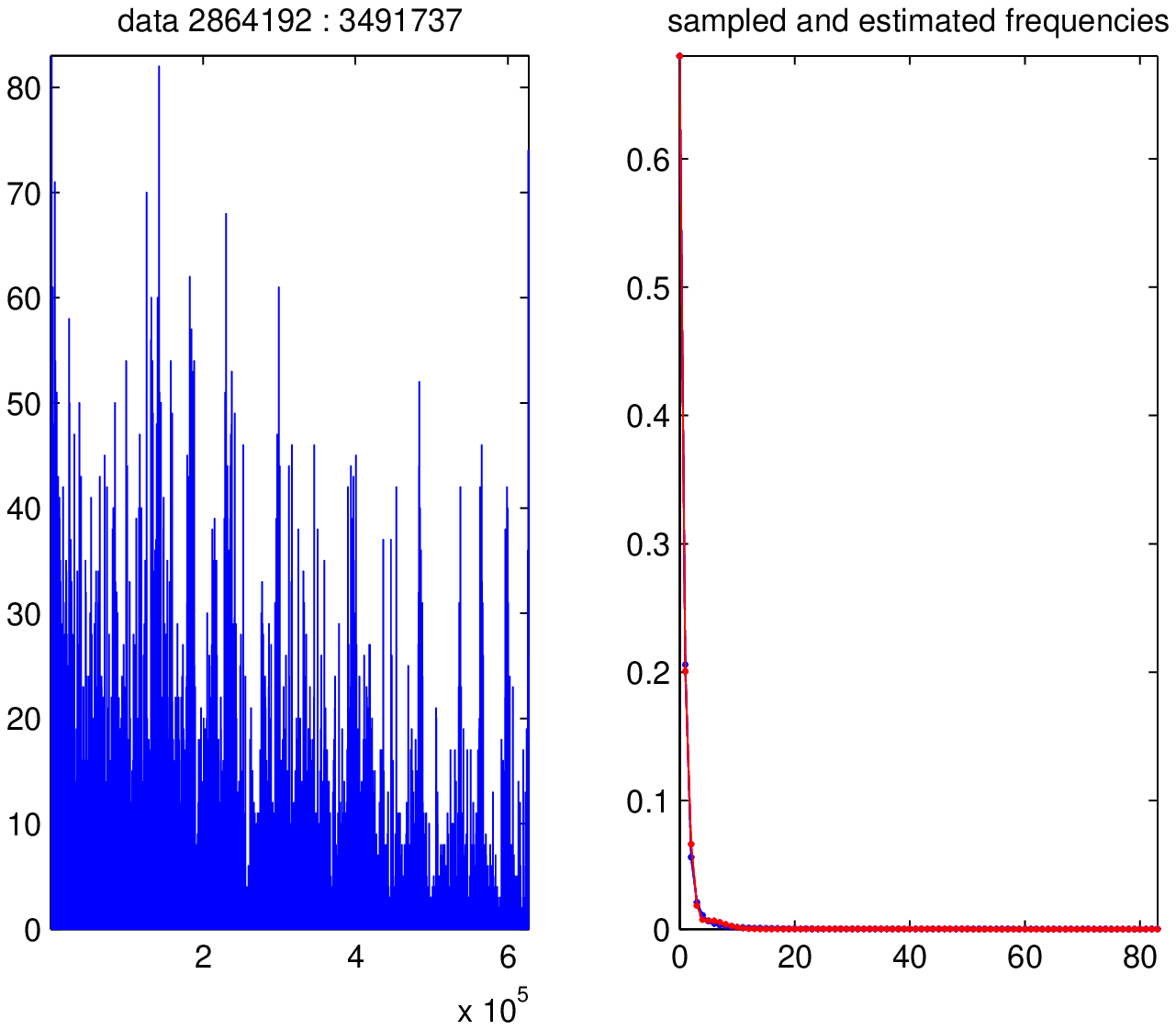} \\
     \includegraphics[width=150mm,height=50mm]{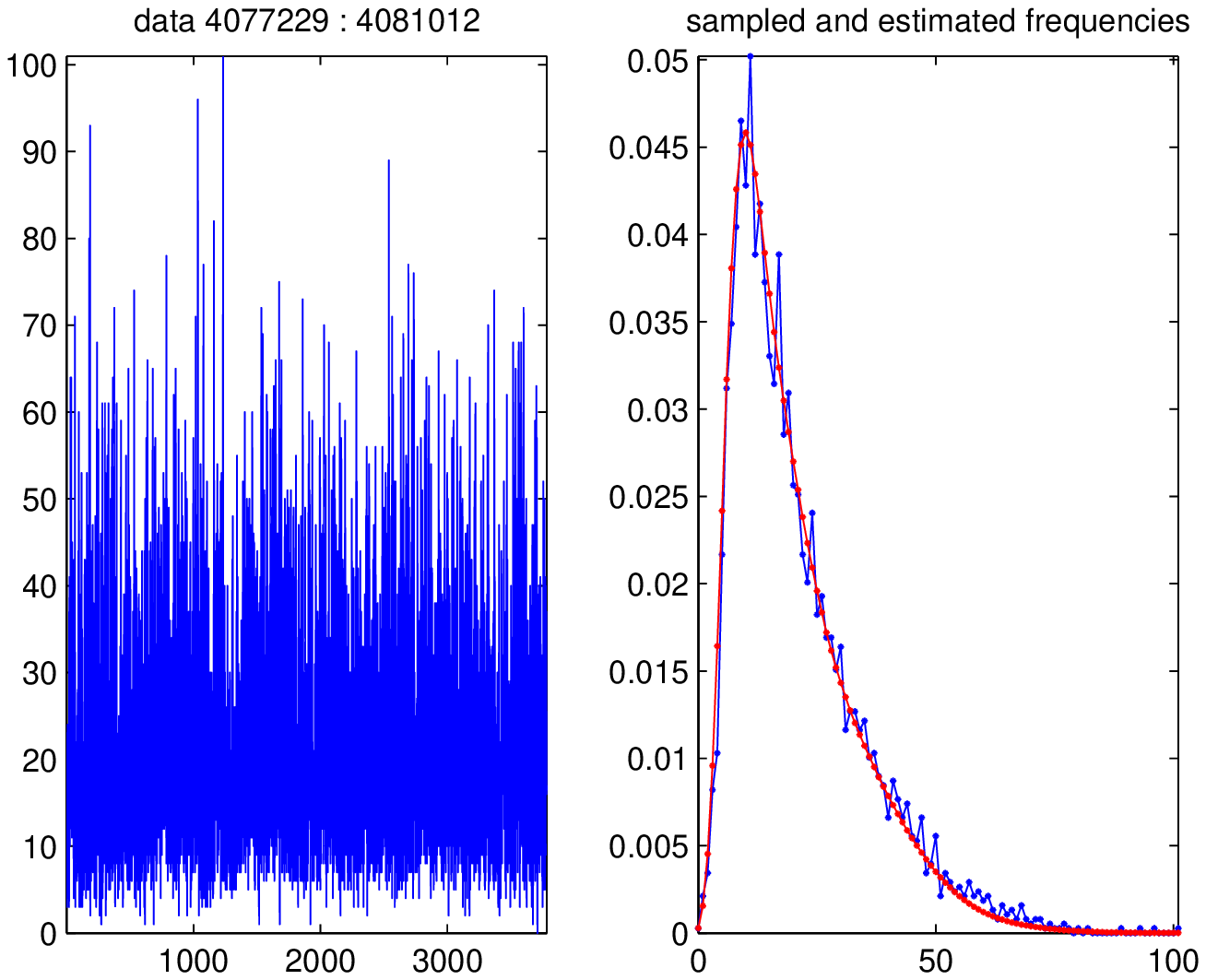}\\
    \end{tabular}
\end{figure}

In order to identify sub-regions of the Saturn rings with distinct properties, 
we  segmented the data  using a method presented in \cite{segmentation} which is designed 
for partitioning of  complicated signals with several non-isolated and oscillating singularities.
In particular, we   applied  the Gabor Continuous Wavelet Transform (see, e.g. \cite{mallat})  to the data  
and  selected the highest scale where the number of  wavelet modulus maxima   takes minimum value. 
%
At this scale, we   segmented  the signal by the method proposed in \cite{segmentation}. We obtained a total of 1531 intervals 
of different sizes.  Figures \ref{fig_real_data_pieces1} and \ref{fig_real_data_pieces2}  refer to six distinct sub-regions of the rings.
The left panels of both figures  show raw data. The right panels exhibit 
the sample and the estimated frequencies, with the penalty parameter obtained by  $DD_{like}$ criterion,
for six  different intervals that are representative of different portions of the data set.

 Note that  in Figure~\ref{fig_real_data_pieces1}, for all three data segments, the estimated parameter $\hat{\pi}_0=0$.
This  is not true for the first and the second panels  of Figure~\ref{fig_real_data_pieces2} 
where $\hat{\pi}_0=0.5059$ and $\hat{\pi}_0=0.2463$, respectively.  
The values of $\Delta_{\nu}$,  defined in \fr{Del_nu}, obtained for the six data segments  are, respectively,  
0.0128,  0.0159,  0.0022,  0.0229,  0.003 and   0.0095, and are consistent with the values obtained in simulations.
Both, the right panels in Figures~\ref{fig_real_data_pieces1}~and~\ref{fig_real_data_pieces2} and the values 
of $\Delta_\nu$, confirm   the ability of the  estimator developed in the paper to accurately explain the 
Saturn's rings data.
%

\section*{Acknowledgments}

Marianna Pensky   was  partially supported by National Science Foundation
(NSF), grant  DMS-1407475. Daniela De Canditiis was entirely supported by the
``Italian Flagship Project Epigenomic'' (http://www.epigen.it/).
The authors   would like to thank Dr. Joshua Colwell for helpful discussions and for providing the data.
The authors  also would  like to thank SAMSI for providing support 
which allowed the author's participation in the 2013-14 LDHD program which was instrumental for 
writing this paper.


 \section {Proofs}
 \label{sec:proofs}
 \setcounter{equation}{0}

Proofs of Theorems~\ref{th:slow_Lasso} and \ref{th:fast_Lasso} are based on the following statement  
which is the trivial modification of Lemma 3 of Pensky (2015).

\begin{lemma} \label{lem:weighted_Lasso}
{\bf (Pensky (2015)). } Let $f$ be the true function and $f_{\bte}$ be its projection 
onto the linear span of the dictionary $\calL_{\calP}$.
Let $\bUp$ be a diagonal matrix with components $\sig_j$, $j=1, \cdots, p$. 
Consider solution of the weighted Lasso problem 
\be \label{las_sol1}
\widehat{\bte} = \arg\min_{\bt}    \lfi   \bt^T \bW  \bW^T \bt -  2  \bt^T \widehat{\bbe}  + 
\alpha \| \bUp \bt \|_1   \rfi. 
\ee
 with  $\bPhi = \bW^T \bW$, $\bobeta = \bPhi \bte$
and   
\be \label{bobeta_cond}
\widehat{\bbe} = \bbe + \sqrt{\eps}\bUp \boeta + \bh, \quad \boeta, \bh \in \RR^p,
\ee
where $\bh$ is a nonrandom vector, $\EE \boeta =0$ and   components $\eta_j$ of vector $\boeta$ are   random variables  
such that, for some $K>0$ and any $\tau>0$, there is a set 
 \be \label{eq:large_dev} 
\Omega = \lfi \omega: \max_{1 \leq j \leq p}  |\eta_j|  \leq K \sqrt{(\tau +1) \log p} \rfi \quad \mbox{with}
\quad \PP(\Omega) \geq 1 - 2 p^{- \tau}.
\ee 
Denote
\be \label{ChCalp}
C_h = \max_{1 \leq j \leq p} \lkv \frac{|h_j|}{ \sig_j\, \sqrt{\eps\, \log p}}\rkv, \quad \quad
C_{\alpha}  = K \sqrt{\tau +1} + C_h.
\ee
If $\alpha_0 = C_{\alpha} \sqrt{\eps \log p}$, then for any $\tau >0$  
and any $\alpha \geq \alpha_0$, then, with probability at least $1 - 2 p^{-\tau}$,  one has
\be \label{slowlaslem}
\| f_{\widehat{\bte}} - f \|_2^2 \leq \inf_{\bt } \lkv  \| f_{\bt } - f \|_2^2 + 4 \alpha \| \bUp \bt \|_1 \rkv.
\ee
Moreover, if  matrices $\bPhi$ and $\bUp$ are such that for some $\mu >1$ and any $J \subset \calP$
\be \label{comp}
\kappa^2 (\mu, J) = \min \lfi \bd \in \calD(\mu, J),\, \| \bd \|_2 \neq 0: \quad 
\frac{\bd^T \bPhi \bd \cdot \Tr(\bUp_J^2)}{\|(\bUp \bd)_{J}\|_1^2} \rfi  >0,
\ee 
and   $\alpha = \varpi \alpha_0$ where $ \varpi \geq  (\mu +1)/(\mu -1)$, then  
for any $\tau >0$   with probability at least $1 - 2 p^{-\tau}$,  one has
\be  \label{fasrlaslem}
\| f_{\widehat{\bte}} - f \|_2^2   \leq   \inf_{\bt, J \subseteq \calP} \lkv  \| f_{\bt } - f \|_2^2 + 4 \alpha  \| (\bUp \bt)_{\Jc} \|_1
+   \frac{(1 + \varpi)^2  C_{\alpha}^2}{\kappa^2 (\mu, J)}  \eps \log p \  \sum_{j \in J} \sig_j^2 \rkv.  
\ee 
\end{lemma}

\medskip

{\bf Proof of Theorem \ref{th:slow_Lasso}. }\  Let    vectors $\bb$ and $\bxi$, 
respectively, have components $b_k = \langle \phi_k, f \rangle$ and $\xi_k$ defined in \fr{z_xi}.
It is easy to see that 
\be \label{for1}
\xi_k - b_k = \frac{1}{n} \sum_{i=1}^n [\psi_k (Y_i) - \EE \psi_k (Y_i)] + H_k
\quad \mbox{with} \quad H_k =  \EE \tpsi_{k, \widehat{\zeta_k}} - b_k
 \ee 
Applying Bernstein inequality, for any $x>0$, obtain
$$
\PP  \lkr \left| n^{-1}\, \sum_{i=1}^n [\psi_k (Y_i) - \EE \psi_k (Y_i)] \right| \geq \frac{x \sig_k}{\sqrt{n}}\rkr 
\leq 2 \exp \lkr -  x^2\,  \lkv 2 + \frac{4 x \sig_k \|\psi_k\|_\infty}{3 \sqrt{n}}\rkv^{-1} \rkr.
$$
Using the fact that $A/(B+C)  \geq \min(A/(2B), A/(2C))$ for any $A,B,C >0$, under condition $n \geq N_0$,
derive
\be \label{for2}
\PP  \lkr  \left| n^{-1}\, \sum_{i=1}^n [\psi_k (Y_i) - \EE \psi_k (Y_i)] \right| \geq x n^{-1/2}\, \sig_k \rkr 
\leq 2 \exp - (x^2/4).
\ee 
Choosing $x= 2 \sqrt{(\tau +1) \log p}$ and recalling that, according to \fr{hatzeta}, $|H_k| = n^{-1/2} \sig_k$,
gather that 
$\PP(|\xi_k - b_k| > n^{-1/2}\, \sig_k  [2 \sqrt{(\tau +1) \log p} +1]) \leq p^{-(\tau +1)}$,
so that
\be \label{large_dev}
\Omega_1 = \lfi \omega:\  \max_{1 \leq k \leq p}\, \lkv \frac{|\xi_k - b_k|}{\sig _k} \rkv 
\leq \frac{2   \sqrt{(\tau +1) \log p} +1}{\sqrt{n}} \rfi
\quad \mbox{with} \quad  \PP(\Omega_1) > 1 - 2 p^{-\tau}.
\ee
Then, validity of Theorem \ref{th:slow_Lasso} follows directly from Lemma \ref{lem:weighted_Lasso} 
with $\eta_k = \xi_k/\sig_k$ and $K = 2$.
\\

\medskip


\noindent
{\bf Proof of Theorem \ref{th:fast_Lasso}. }  Validity of inequality \fr{f_error} follows from \fr{large_dev}
and Lemma 1 with $K = 2$.

In order to establish upper bounds for $(\hat{\pi}_0 - \pi_0)^2$, note that
$\PP(Y=0) = \pi_0 + \bte^T \bu$, so that
\be \label{pi0_dif}
 \hat{\pi}_0 - \pi_0  = \lfi
\begin{array}{ll}
\nu_0 - \PP(Y=0) - \bd^T \bu, & \mbox{if}\ \ \nu_0 - \widehat{\bte}^T \bu \geq 0,\\
- \pi_0,  & \mbox{if}\ \ \nu_0 - \widehat{\bte}^T \bu < 0,
\end{array} \right.
\ee 
where $\bd = \widehat{\bte} - \bte_0$. Then, by standard arguments 
(see, e.g. Dalalyan {\it  et al.} (2014)), one has 
$$
\bd^T \bPhi \bd \leq \bd^T(\bxi - \bb) + \alpha(\|\bUp \bte_0\|_1 - \|\bUp \hat{\bte}\|_1)
$$
For $\omega \in \Omega_1$ where $\Omega_1$ is defined in \fr{large_dev}, one has 
\be \label{ineq1}
\bd^T \bPhi \bd + (\alpha - \alpha_0) \| (\bUp \bd)_{J_0^c}\|_1 \leq (\alpha + \alpha_0) \| (\bUp \bd)_{J_0}\|_1.
\ee
Therefore, $\bd \in \calD (\mu, J_0)$ where $\calD (\mu, J)$ is defined in \fr{Dset}, and 
due to $J_0  \in \calG (C_{\sig})$,  the following inequality holds: 
\be \label{ineq2}
\| \bd_{J_0^c}\|_1 \leq \mu C_{\sig}  \|\bd_{J_0}\|_1.
\ee
Hence, by  \fr{vartsm}, $\bd^T \bPhi \bd \geq \vartheta (s,m, \mu, C_\sig)$. 
On the other hand, inequality \fr{ineq1} yields
$\bd^T \bPhi \bd \leq (\alpha + \alpha_0) \|\bd_{J_0}\|_2 \sqrt{\Tr(\bUp_{J_0}^2)}$,
so that 
\be \label{ineq3}
\|\bd_{J_0}\|_2 \leq \frac{(\alpha + \alpha_0)}{\vartheta (s,m, \mu, C_\sig)} \ \sqrt{\Tr(\bUp_{J_0}^2)}.
\ee
Using \fr{ineq2} and \fr{ineq3}, obtain that, for any $\omega \in \Omega_1$,
\beqns  
|\bd^T \bu| & \leq & \|\bd_{J_0}\|_1 \|\bu_{J_0}\|_\infty + \|\bd_{J_0^c}\|_1 \|\bu_{J_0^c}\|_\infty 
\leq \|\bd_{J_0}\|_1 \|\bu\|_\infty (1 + \mu C_\sig) \\
& \leq & \sqrt{s} \|\bd_{J_0}\|_2  \|\bu\|_\infty (1 + \mu C_\sig) \leq 
 \sqrt{s}\, \alpha_0  \, (1 + \varpi)  \|\bu\|_\infty (1 + \mu C_\sig) \ \sqrt{\Tr(\bUp_{J_0}^2)}/\vartheta (s,m, \mu, C_\sig).
\eeqns
In addition, there exists a set  $\Omega_2$ such that, for $\omega \in \Omega_2$, one has 
$|\nu_0 - \PP(Y=0)| \leq \sqrt{(\tau \log p)/n}$ and $\PP(\Omega_2) \geq 1 - 2 p^{-\tau}$.
Let $\Omega = \Omega_1 \cap \Omega_2$. Then, $\PP(\Omega) \geq 1 - 4 p^{-\tau}$ and, for $\omega \in \Omega$,
\be \label{ineq4}
|\nu_0 - \PP(Y=0) - \bd^T \bu| \leq \frac{\sqrt{s}\, \alpha_0  \, \sqrt{\Tr(\bUp^2_{J_0})}\, 
(1 + \varpi)\, \|\bu\|_\infty (1 + \mu C_\sig)}{\vartheta (s,m, \mu, C_\sig)}\, 
   + \frac{\sqrt{\tau \log p}}{\sqrt{n}}.
\ee
Inequality \fr{ineq4} provides an upper bound for the error if $\nu_0 - \widehat{\bte}^T \bu \geq 0$.
If $\nu_0 - \widehat{\bte}^T \bu < 0$, note that 
$$
0 \leq \pi_0 = (\PP(Y=0) - \nu_0 - \bd^T \bu) + (\nu_0 - \widehat{\bte}^T \bu) \leq 
\PP(Y=0) - \nu_0 - \bd^T \bu,
$$
and again apply  \fr{ineq4}.
Finally, plugging in the value  of $\alpha_0$ and using inequality for $H_0$, derive that, for $\omega \in \Omega$,
inequality \fr{pi_error} holds.


\newpage

\noindent
\begin{table}
\caption{Average values of $\Delta_g$ (with their standard deviations in parentheses) 
over 100 simulation runs with  $n=10000$ }
\label{tab_res1}
\begin{center}\scriptsize
\begin{tabular}{ c | c c c c }
 test case & $OPT$ &  $DD_{l2}$  & $DD_{like}$ &  $NDE$     \\
\hline
case1 & 0.0007  (0.0010)    & 0.0023  (0.0028)    & 0.0022  (0.0030)     & 0.1183  (0.8307)    \\
case2 & 0.0471  (0.0197)    & 0.2214  (0.0503)    & 0.0507  (0.0305)     & 0.0613  (0.0716)    \\ 
case3 & 0.0142  (0.0398)    & 0.0191  (0.0127)    & 0.0138  (0.0087)     & 0.0190  (0.0399)    \\ 
case4 & 0.0043  (0.0021)    & 0.0054  (0.0032)    & 0.0061  (0.0052)     & 0.0298  (0.0657)    \\ 
case5 & 0.0042  (0.0033)    & 0.0023  (0.0029)    & 0.0014  (0.0021)     & 1.0000  (0.0000)    \\ 
case6 & 0.0793  (0.0247)    & 0.4318  (0.0554)    & 0.0839  (0.0241)     & 0.3383  (0.0085)    \\ 
case7 & 0.0067  (0.0012)    & 0.0009  (0.0008)    & 0.0008  (0.0008)     & -    \\ 
case8 & 0.0060  (0.0010)    & 0.0068  (0.0009)    & 0.0069  (0.0010)     & -    \\ 
case9 & 0.0085  (0.0013)    & 0.0099  (0.0014)    & 0.0111  (0.0026)     & -   \\ 
 \end{tabular}
\end{center}
\end{table}

\noindent
\begin{table}
\caption{Average values of $\Delta_\nu$ (with their standard deviations in parentheses) 
over 100 simulation runs with $n=10000$ }
\label{tab_res2}
\begin{center}\scriptsize
\begin{tabular}{ c | c c c  c }
 test case & $OPT$ &  $DD_{l2}$  & $DD_{like}$  &  $NDE$     \\
\hline
case1 & 0.0011  (0.0009)    & 0.0006  (0.0005)    & 0.0007  (0.0005)      & 0.0675  (0.5470)    \\ 
case2 & 0.0013  (0.0007)    & 0.0088  (0.0024)    & 0.0014  (0.0010)      & 0.0228  (0.0431)    \\ 
case3 & 0.0002  (0.0001)    & 0.0006  (0.0005)    & 0.0003  (0.0003)      & 0.1130  (0.0653)    \\ 
case4 & 0.0014  (0.0009)    & 0.0011  (0.0006)    & 0.0012  (0.0009)      & 0.0205  (0.0530)    \\ 
case5 & 0.0045  (0.0010)    & 0.0043  (0.0010)    & 0.0044  (0.0009)      & 1.0000  (0.0000)    \\ 
case6 & 0.0465  (0.1402)    & 0.0288  (0.0045)    & 0.0041  (0.0020)      & 0.4376  (0.0618)    \\ 
case7 & 0.0013  (0.0002)    & 0.0006  (0.0002)    & 0.0006  (0.0002)      & -    \\ 
case8 & 0.0039  (0.0009)    & 0.0018  (0.0004)    & 0.0018  (0.0004)      & -    \\ 
case9 & 0.0035  (0.0006)    & 0.0019  (0.0007)    & 0.0020  (0.0006)      & -    \\ 
 \end{tabular}
\end{center}
\end{table}
\noindent
\begin{table}
\caption{Average values of $\Delta_g$ (with their standard deviations in parentheses) 
over 100 simulation runs with $n=5000$  }
\label{tab_res3}
\begin{center}\scriptsize
\begin{tabular}{ c | c c c  c }
 test case & $OPT$ &  $DD_{l2}$  & $DD_{like}$  &  $NDE$     \\
\hline
case1 & 0.0006  (0.0008)    & 0.0038  (0.0049)    & 0.0030  (0.0046)     & 0.2424  (1.5002)    \\ 
case2 & 0.0590  (0.0343)    & 0.2106  (0.0549)    & 0.0640  (0.0428)     & 0.2048  (0.2196)    \\
case3 & 0.0148  (0.0097)    & 0.0251  (0.0310)    & 0.0178  (0.0119)     & 0.0309  (0.0650)    \\ 
case4 & 0.0055  (0.0019)    & 0.0074  (0.0051)    & 0.0086  (0.0060)     & 0.0493  (0.1123)    \\ 
case5 & 0.0069  (0.0052)    & 0.0044  (0.0048)    & 0.0024  (0.0036)     & 1.0000  (0.0000)    \\ 
case6 & 0.0830  (0.0277)    & 0.4068  (0.0856)    & 0.0879  (0.0266)    & 0.3456  (0.0110)    \\ 
case7 & 0.0077  (0.0035)    & 0.0023  (0.0030)    & 0.0023  (0.0030)    & -    \\ 
case8 & 0.0065  (0.0021)    & 0.0074  (0.0020)    & 0.0075  (0.0021)    & -    \\ 
case9 & 0.0096  (0.0023)    & 0.0114  (0.0023)    & 0.0128  (0.0029)     & -    \\ 
 \end{tabular}
\end{center}
\end{table}

\noindent
\begin{table}
\caption{Average values of $\Delta_\nu$ (with their standard deviations in parentheses) 
over 100 simulation runs with $n=5000$  }
\label{tab_res4}
\begin{center}\scriptsize
\begin{tabular}{ c | c c c c }
 test case & $OPT$ &  $DD_{L2}$  & $DD_{like}$ &  $NDE$     \\
\hline
case1 & 0.0017  (0.0016)    & 0.0010  (0.0007)    & 0.0012  (0.0007)      & 0.1393  (1.0084)    \\
case2 & 0.0020  (0.0013)    & 0.0090  (0.0029)    & 0.0022  (0.0014)      & 0.1184  (0.1475)    \\ 
case3 & 0.0004  (0.0002)    & 0.0009  (0.0013)    & 0.0006  (0.0004)      & 0.1131  (0.0948)    \\ 
case4 & 0.0022  (0.0014)    & 0.0016  (0.0008)    & 0.0018  (0.0013)      & 0.0346  (0.0859)    \\ 
case5 & 0.0087  (0.0019)    & 0.0084  (0.0018)    & 0.0085  (0.0018)      & 1.0000  (0.0000)    \\ 
case6 & 0.0377  (0.1361)    & 0.0285  (0.0068)    & 0.0057  (0.0030)      & 0.4608  (0.0849)    \\ 
case7 & 0.0020  (0.0003)    & 0.0013  (0.0003)    & 0.0013  (0.0003)      & -    \\ 
case8 & 0.0052  (0.0011)    & 0.0032  (0.0008)    & 0.0032  (0.0008)      & -    \\ 
case9 & 0.0044  (0.0010)    & 0.0031  (0.0010)    & 0.0031  (0.0009)      & -    \\ 
 \end{tabular}
\end{center}
\end{table}

\noindent
\begin{table}
\caption{Average values of $\Delta_g$ (with their standard deviations in parentheses) 
over 100 simulation runs with $n=1000$  }
\label{tab_res5}
\begin{center}\scriptsize
\begin{tabular}{ c | c c c c }
 test case & $OPT$ &  $DD_{l2}$  & $DD_{like}$  &  $NDE$     \\
\hline
case1 & 0.0040  (0.0097)    & 0.0221  (0.0258)    & 0.0176  (0.0207)     & 0.3004  (0.9331)    \\ 
case2 & 0.0992  (0.0760)    & 0.1973  (0.0718)    & 0.1335  (0.0983)     & 0.5370  (0.0960)    \\ 
case3 & 0.0533  (0.0889)    & 0.0753  (0.0838)    & 0.0662  (0.0894)     & 0.1127  (0.3912)    \\ 
case4 & 0.0069  (0.0014)    & 0.0178  (0.0183)    & 0.0179  (0.0135)     & 0.1393  (0.3381)    \\ 
case5 & 0.0170  (0.0108)    & 0.0217  (0.0253)    & 0.0152  (0.0223)     & 1.0000  (0.0000)    \\ 
case6 & 0.1223  (0.0710)    & 0.3151  (0.1409)    & 0.1270  (0.0759)     & 0.4479  (0.2572)    \\ 
case7 & 0.0133  (0.0115)    & 0.0102  (0.0118)    & 0.0098  (0.0118)     & -    \\ 
case8 & 0.0142  (0.0137)    & 0.0156  (0.0139)    & 0.0156  (0.0154)     & -    \\ 
case9 & 0.0121  (0.0073)    & 0.0163  (0.0110)    & 0.0160  (0.0101)     & -    \\ 
 \end{tabular}
\end{center}
\end{table}

\noindent
\begin{table}
\caption{Average values of $\Delta_\nu$ (with their standard deviations in parentheses) 
over 100 simulation runs with $n=1000$  }
\label{tab_res6}
\begin{center}\scriptsize
\begin{tabular}{ c | c c c  c }
 test case & $OPT$ &  $DD_{l2}$  & $DD_{like}$ &  $NDE$     \\
\hline
case1 & 0.0063  (0.0042)    & 0.0043  (0.0026)    & 0.0047  (0.0027)    & 0.1458  (0.5377)    \\ 
case2 & 0.0076  (0.0051)    & 0.0117  (0.0047)    & 0.0084  (0.0048)    & 0.3498  (0.0937)    \\ 
case3 & 0.0021  (0.0022)    & 0.0031  (0.0033)    & 0.0027  (0.0037)    & 0.2149  (0.3773)    \\ 
case4 & 0.0075  (0.0068)    & 0.0046  (0.0029)    & 0.0048  (0.0032)    & 0.0967  (0.3578)    \\ 
case5 & 0.0427  (0.0091)    & 0.0411  (0.0090)    & 0.0416  (0.0090)    & 1.0000  (0.0000)    \\ 
case6 & 0.0157  (0.0044)    & 0.0304  (0.0127)    & 0.0166  (0.0067)    & 0.5344  (0.1001)    \\ 
case7 & 0.0072  (0.0018)    & 0.0067  (0.0018)    & 0.0067  (0.0018)    & -    \\ 
case8 & 0.0154  (0.0038)    & 0.0143  (0.0038)    & 0.0142  (0.0037)    & -   \\ 
case9 & 0.0120  (0.0034)    & 0.0111  (0.0032)    & 0.0111  (0.0032)    & -    \\ 
 \end{tabular}
\end{center}
\end{table}
%



\end{document}